\documentclass[12pt,twocolumn]{emulateapj}

\usepackage{color}
\usepackage[usenames,dvipsnames]{xcolor}
\usepackage[colorlinks,urlcolor=blue,citecolor=blue,linkcolor=blue]{hyperref}
\usepackage[multidot]{grffile} 
\usepackage{amsmath}
\usepackage{array}

\slugcomment{}

\begin{document} 

\title{Target Selection for the SDSS-IV APOGEE-2 Survey} 
\shorttitle{APOGEE-2 Targeting}

\author{
G.~Zasowski\altaffilmark{1,2}, 
R.~E.~Cohen\altaffilmark{2},
S.~D.~Chojnowski\altaffilmark{3},
F.~Santana\altaffilmark{4},
R.~J.~Oelkers\altaffilmark{5}, \\
B.~Andrews\altaffilmark{6},
R.~L.~Beaton\altaffilmark{7}, 
C.~Bender\altaffilmark{8}, 
J.~C.~Bird\altaffilmark{5}, 
J.~Bovy\altaffilmark{9},
J.~K.~Carlberg\altaffilmark{2}, 
K.~Covey\altaffilmark{10}, 
K.~Cunha\altaffilmark{8,11},
F.~Dell'Agli\altaffilmark{12},
Scott~W.~Fleming\altaffilmark{2},
P.~M.~Frinchaboy\altaffilmark{13}, 
D.~A.~Garc\'{i}a-Hern\'{a}ndez\altaffilmark{12},
P.~Harding\altaffilmark{14},
J.~Holtzman\altaffilmark{3},
J.~A.~Johnson\altaffilmark{15}, 
J.~A.~Kollmeier\altaffilmark{7}, 
S.~R.~Majewski\altaffilmark{16}, 
Sz.~M{\'e}sz{\'a}ros\altaffilmark{17,18},
J.~Munn\altaffilmark{19},
R.~R.~Mu\~{n}oz\altaffilmark{4},
M.~K.~Ness\altaffilmark{20}, 
D.~L.~Nidever\altaffilmark{21}, 
R.~Poleski\altaffilmark{15,22},
C.~Rom\'{a}n~Z\'{u}\~{n}iga\altaffilmark{23}, 
M.~Shetrone\altaffilmark{24}, 
J.~D.~Simon\altaffilmark{7}, 
V.~V.~Smith\altaffilmark{21},
J.~S.~Sobeck\altaffilmark{25}, 
G.~S.~Stringfellow\altaffilmark{26},
L.~Szigeti{\'a}ros\altaffilmark{17},
J.~Tayar\altaffilmark{15}, 
N.~Troup\altaffilmark{16,27} 
}
\shortauthors{Zasowski et al.}

\altaffiltext{1}{Department of Physics \& Astronomy, University of Utah, Salt Lake City, UT, 84112, USA; gail.zasowski@gmail.com}
\altaffiltext{2}{Space Telescope Science Institute, Baltimore, MD, 21218, USA}
\altaffiltext{3}{Department of Astronomy, New Mexico State University, Las Cruces, NM, 88001, USA}
\altaffiltext{4}{Departamento de Astronom\'ia, Universidad de Chile, Santiago, Chile}
\altaffiltext{5}{Department of Physics \& Astronomy, Vanderbilt University, Nashville, TN, 37235, USA}
\altaffiltext{6}{PITT PACC, Department of Physics \& Astronomy, University of Pittsburgh, Pittsburgh, PA, 15260, USA}
\altaffiltext{7}{The Observatories of the Carnegie Institution for Science, Pasadena, CA 91101, USA}
\altaffiltext{8}{Steward Observatory, The University of Arizona, Tucson, AZ, 85719, USA}
\altaffiltext{9}{Department of Astronomy and Astrophysics \& Dunlap Institute for Astronomy and Astrophysics, University of Toronto, Toronto, Ontario M5S 3H4, Canada}
\altaffiltext{10}{Department of Physics \& Astronomy, Western Washington University, Bellingham, WA, 98225, USA}
\altaffiltext{11}{Observat\'{o}rio Nacional, 20921-400 So Crist\'{o}vao, Rio de Janeiro, RJ, Brazil}
\altaffiltext{12}{Departamento de Astrof\'{\i}sica, Universidad de La Laguna, \& Instituto de Astrof\'{\i}sica de Canarias, La Laguna, Tenerife, Spain }
\altaffiltext{13}{Department of Physics \& Astronomy, Texas Christian University, Fort Worth, TX, 76129, USA}
\altaffiltext{14}{Department of Astronomy, Case Western Reserve University, Cleveland, OH, 44106, USA}
\altaffiltext{15}{Department of Astronomy, The Ohio State University, Columbus, OH, 43210, USA}
\altaffiltext{16}{Department of Astronomy, University of Virginia, Charlottesville, VA, 22903, USA}
\altaffiltext{17}{ELTE E\"otv\"os Lor\'and University, Gothard Astrophysical Observatory, Szombathely, Hungary}
\altaffiltext{18}{Premium Postdoctoral Fellow of the Hungarian Academy of Sciences}
\altaffiltext{19}{US Naval Observatory, Flagstaff Station, Flagstaff, AZ, 86005, USA}
\altaffiltext{20}{Max-Planck-Institut f\"{u}r Astronomie, D-69117, Heidelberg, Germany}
\altaffiltext{21}{National Optical Astronomy Observatory, Tucson, AZ, 85719, USA}
\altaffiltext{22}{Warsaw University Observatory, 00-478 Warszawa, Poland}
\altaffiltext{23}{Instituto de Astronom\'{i}a, Universidad Nacional Aut\'{o}noma de M\'{e}xico, Ensenada, BC, 22860, Mexico}
\altaffiltext{24}{McDonald Observatory, University of Texas at Austin, TX, 79734, USA}
\altaffiltext{25}{Department of Astronomy, University of Washington, Seattle, WA, 98195, USA}
\altaffiltext{26}{Center for Astrophysics and Space Astronomy, Department of Astrophysical and Planetary Sciences, University of Colorado, Boulder, CO, 80309, USA}
\altaffiltext{27}{Department of Physics, Salisbury University, Salisbury, MD, 21801, USA}
\setcounter{footnote}{0}

\begin{abstract}
APOGEE-2 is a high-resolution, near-infrared spectroscopic survey observing $\sim$3$\times$10$^5$ stars across the entire sky.  It is the successor to APOGEE and is part of the Sloan Digital Sky Survey IV (SDSS-IV).  APOGEE-2 is expanding upon APOGEE's goals of addressing critical questions of stellar astrophysics, stellar populations, and Galactic chemodynamical evolution using (1) an enhanced set of target types and (2) a second spectrograph at Las Campanas Observatory in Chile.  APOGEE-2 is targeting red giant branch (RGB) and red clump (RC) stars, RR Lyrae, low-mass dwarf stars, young stellar objects, and numerous other Milky Way and Local Group sources across the entire sky from both hemispheres.  In this paper, we describe the APOGEE-2 observational design, target selection catalogs and algorithms, and the targeting-related documentation included in the SDSS data releases.
\end{abstract}



\section{Introduction} \label{sec:intro}
\setcounter{footnote}{0}

Understanding the parameter space of galaxies, and the processes that drive galaxy evolution, require precise, accurate, multi-wavelength measurements of not only galaxies, but also of their stellar and interstellar building blocks.  Photometric and spectroscopic surveys of galaxies can provide global dynamical and chemical properties as a function of redshift, along with spatial variations in these properties down to scales of $\sim$1~kpc for galaxies at $z \sim 0.1$.  This is a factor of several larger than the resolution achievable with current MHD simulations of $L_*$ galaxies \citep[$\sim$1--50~pc; e.g.,][]{Wetzel_2016_Latte,Hopkins_2017_FIRE2}.

In the Local Group, individual stars can be targeted for spectroscopy in several satellite galaxies, including the Magellanic Clouds, and in the haloes of M31 and M33.  But our only access to large numbers of stars in the disk and bulge of a typical $L_*$-sized galaxy (where most of the stars in the Universe reside) comes from studies of the Milky Way (MW) Galaxy.  A comprehensive understanding of these stellar populations, and the interstellar medium (ISM) between them, provides a critical {\it and unique} data point to compare to the end products of cosmological and galactic evolutionary models.  This is the advantage of the so-called ``near field cosmology'', the use of high-resolution information at very low redshift to constrain the generic physical processes operating at very high redshift.

Surveys of the MW's stars make up the oldest dedicated astronomical efforts, dating back to Hipparcos and \citet{Herschel_1785_Oheavens}, among others.  Starting in the mid 20th century \citep[shortly after it was proven that the MW {\it was} a galaxy\footnote{We note that numerous earlier astronomers, including Nas\={\i}r al-D\={\i}n T\={u}s\={\i} and Immanuel Kant, speculated that the Galaxy was a distinct body composed of clustered stars, but technology did not enable the scientific proof of these conjectures until the late 19th and early 20th centuries.}; e.g.,][]{Curtis_1917_islanduniverses,Hubble_1926_extragalacticnebulae}, photometric stellar surveys began to reveal complex structures in the MW, including numerous streams in the halo and the presence of two, seemingly coplanar stellar disks.  These surveys have been critically discriminatory datasets used to fuel or reject theories of galaxy formation and evolution.  While the {\it basic} structure and stellar populations in our Galaxy have been known for this past century, surprises do continue to this day, for example, with the discoveries of the X-shaped bulge in the center \citep{McWilliam_10_xshapedbulge,Nataf_2010_dualbulgeRC,Ness_2016_Xshapedbulge} and the presence of multiple stellar populations in globular clusters \citep[e.g.,][]{Piotto_2009_GCmultiplepops,Gratton_2012_multipleGCpops}.  

The pace of discovery has been further spurred by the accessibility of long-wavelength optical and infrared (IR) detection technology, which allow photometric and spectroscopic measurements of stars behind the previously-inpenetrable dust in the Galactic disk.  These datasets have provided a wealth of both refinements to existing knowledge and new discoveries: for example, 
the properties of stellar substructure in the outskirts of the disk \citep[with 2MASS;][]{Rocha-Pinto_2003_GASS},
the existence of a long bar extending 2~kpc beyond the bulge \citep[with GLIMPSE;][]{Benjamin_05_glimpse}, 
variations in the IR dust exinction law \citep[with 2MASS and GLIMPSE;][]{Nishi_09_extlaw,Zasowski_09_extlaw}, 
spatially varying peaks in the bulge metallicity distribution \citep[with ARGOS;][]{Ness_2013_argos}, 
and second-generation asymptotic giant branch (AGB) stars and multiple stellar populations in metal-poor and metal-rich bulge
globular clusters, respectively \citep[with APOGEE;][]{GarciaHernandez_2ndGenAGB,Schiavon_2017_APOGEEbulgeGCs}.  

Of course, it is not simply the size or spatial coverage of the observed sample that leads to a deeper understanding of galaxy evolution across cosmic time.  Stars occupy a high dimensionality of phase space ($\vec{X}$, $\vec{V}$, [X/H], etc), and mapping their distribution along all of these axes is a key aim of ``galactic archeology'', an approach named for its parallels to archeological studies of humans.  In both fields, many pieces of complementary evidence from different aspects of a system's evolution are fitted together into a coherent, multi-layered model.  In archeological/historical studies of humans, this may mean combining evidence from tax records, public art, oral histories, and rubbish heaps to understand a complex society's rise and fall.  In galactic archeology, we combine as many dimensions of spatial, dynamical, chemical, and age information as possible to produce a comprehensive model of either a particular galaxy's evolutionary path, or the full range of evolutionary paths that galaxies can have.

Stellar spectroscopy can access dimensions of stellar phase space largely unreachable by photometry: namely, radial velocities (RVs), precise stellar atmospheric parameters\footnote{The ability to derive surface gravities from (photometric) asteroseismic measurements is a relatively recent exception.}, and metallicity information; at high spectral resolution, individual elemental abundances also become available, which reflect the detailed enrichment history of each star's natal ISM cloud.  When this spectroscopy is performed in the IR on luminous giant stars, these dimensions can be readily obtained for stars throughout the dusty, densely packed inner regions of the MW.  

This goal of obtaining high-dimensional stellar information throughout the MW served as the inspiration for the APOGEE survey \citep{Majewski_2017_apogeeoverview}, one of the components of the third generation of the Sloan Digital Sky Survey \citep[SDSS-III;][]{Eisenstein_11_sdss3overview}.  During 2011--2014, APOGEE obtained high-resolution ($R \sim 22,000$), high SNR, $H$-band spectra for $\sim$163,000 stars in the bulge, disk, and halo of the MW \citep{Zasowski_2013_apogeetargeting,Holtzman_2015_apogeedata}, spanning up to $\sim$12~kpc away in the midplane.  Data products publicly released by the survey \citep{Holtzman_2015_apogeedata} include the raw and reduced spectra, radial velocities \citep{Nidever_2015_apogeereduction}, fundamental stellar parameters \citep{GarciaPerez_2016_aspcap,Meszaros_2013_aspcapcalib,Zamora_2015_apogeelibraries}, and abundances for $\sim$20-25 elements per star \citep{Shetrone_2015_apogeelinelist,Smith_2013_apogeelinelistFTS}.

The APOGEE-2 survey \citep{Majewski_2016_apogee1-2}, part of SDSS-IV \citep[2014--2020;][]{Blanton_2017_sdss4}, expands and significantly enhances the original APOGEE sample,  with the key addition of a second spectrograph at Las Campanas Observatory (LCO) to make a unqiue {\it all-sky} {\it H}-band spectroscopic survey.  APOGEE-2 thus comprises two complementary components: APOGEE-2 North (APOGEE-2N), using the Sloan Foundation 2.5-meter telescope at APO \citep{Gunn_2006_sloantelescope} and original APOGEE spectrograph  \citep{Wilson_2012_apogee}, and APOGEE-2 South (APOGEE-2S), using the duPont 2.5-m telescope at LCO and a cloned spectrograph.  The APOGEE-2 sample expands the original red giant sample in both distance and spatial coverage, and further enhancements come from an increasing diversity of targeted objects and scientific goals.  The first public release of APOGEE-2 data is contained in Data Release 14\footnote{\url{http://www.sdss.org/dr14/}} in the summer of 2017 \citep{Abolfathi_2017_SDSSDR14}.

In this paper, we present the target selection and observing strategy of the APOGEE-2 survey.  These are critical factors to understand when analyzing data from {\it any} survey, to account for selection effects and biases in the base sample \citep[e.g.,][]{Schlesinger_2012_SEGUEMDF}.  In \S\ref{sec:fields}, we describe the design of the survey footprint and the organization and prioritization of the targeted objects in each pointing. The logistics of identifying selection criteria applied to individal sample objects is presented in \S\ref{sec:targeting_flags}.  In \S\ref{sec:main_sample}, we outline the criteria used to select the primary red giant sample, and \S\S\ref{sec:clusters}--\ref{sec:ancillary} contain information on the numerous other classes of targets, including stellar clusters (\S\ref{sec:clusters}), satellite galaxies (\S\ref{sec:sat_galaxies}), and ancillary programs (\S\ref{sec:ancillary}).  In \S\ref{sec:calibration_targets}, we describe special calibration targets observed for the purposes of correcting observational artifacts (\S\ref{sec:calibration_observation}) and comparing derived stellar parameters to previous work (\S\S\ref{sec:calibration_params}--\ref{sec:calibration_surveys}).  A summary of the targeting information included in the SDSS data releases is presented in \S\ref{sec:dr_targeting_info}.  A glossary of SDSS- and APOGEE-specific terminology is provided in Appendix~\ref{sec:glossary}.

\section{Field Plan and Observing Strategies} 
\label{sec:fields}

\begin{figure*}
\begin{center}
\includegraphics[angle=0,trim=0in 0in 0in 0in, clip, width=\textwidth]{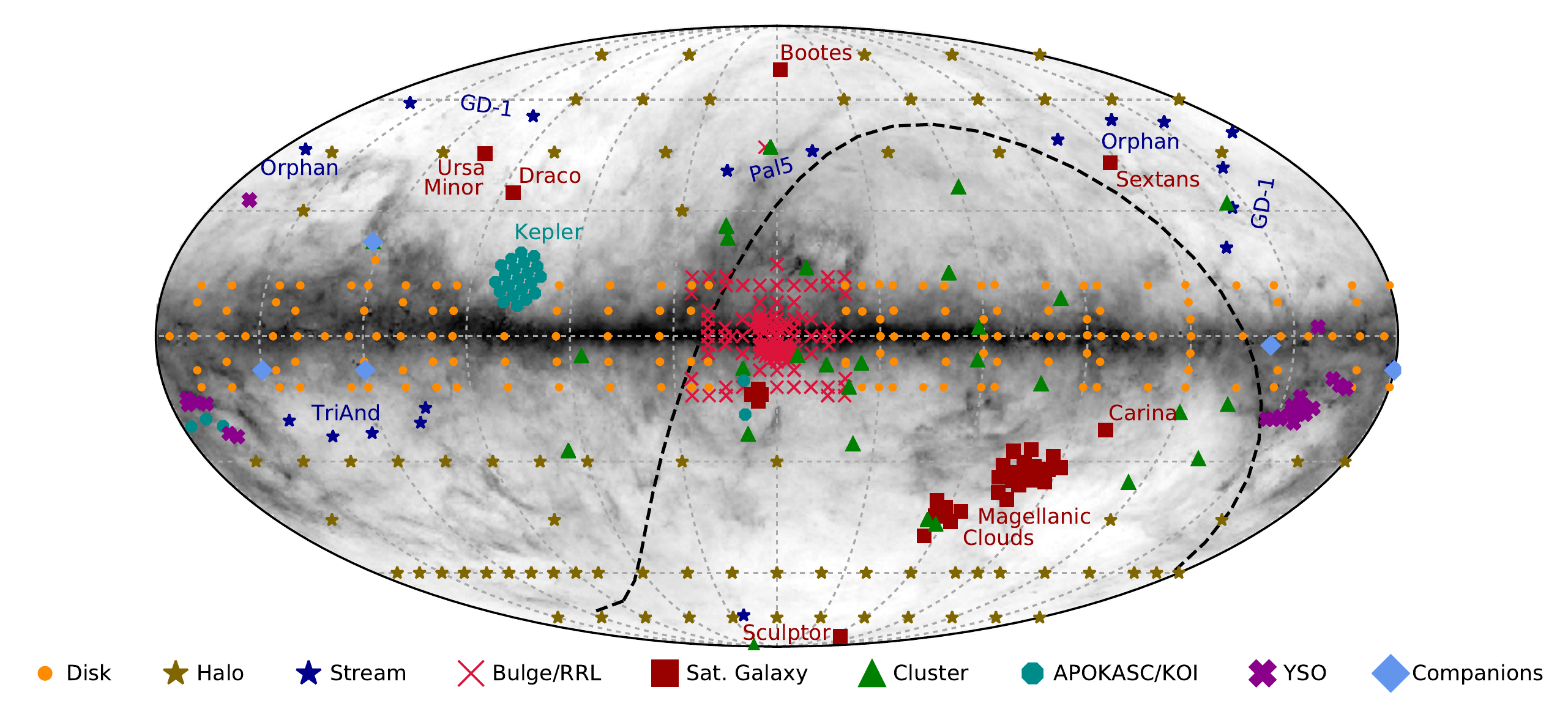} 
\caption{The current APOGEE-2 field plan in Galactic coordinates, with fields colored by the primary driver of their placement.  The background grayscale is the integrated $E(B-V)$ map from \citet{Schlegel_1998_dustmap}; the concentration of APOGEE-2 observations in some of the dustiest parts of the Milky Way highlights the targeting advantage of NIR surveys compared to optical ones.  The dashed line indicates the approximate declination limit of $-10^\circ$ adopted for the Northern (upper, left) and Southern (lower, right)  survey components; fields with declinations within $\sim$15$^\circ$ of this line may be observed from either hemisphere. The gray meridians and parallels are spaced every $\Delta l,\Delta b = 30^\circ$.
}
\label{fig:fieldmap}
\end{center}
\end{figure*}

\subsection{Field Properties}
\label{sec:field_properties}

Throughout this paper, we will refer to ``fields'' or ``pointings'',
which indicates a patch of sky spanned by a given set or design of targets (\S\ref{sec:design_defs}), defined by a central position and radial extent (see below).  
We will often classify these pointings as ``disk fields'', ``bulge fields'', ``ancillary fields'', etc., 
which indicates either the dominant MW component in the field or the motivation behind targeting that particular patch of sky.
The current APOGEE-2 field plan is shown in Figure~\ref{fig:fieldmap}, with fields colored by type.

The radial extent of a field is driven by the field of view (FOV) of the telescope and the declination of the field itself.  The 2.5-m SDSS telescope at APO has a maximal FOV of 1.5$^\circ$ in radius, and the APOGEE instrument on the 2.5-m duPont telescope at LCO has a maximal FOV of 0.95$^\circ$ in radius; these are the maximum design radii of APOGEE-2N and -2S fields, respectively.  However, smaller size limits may be imposed on fields observed at high airmass, due to the high differential refraction across the large FOV.  Thus, for example, fields with $\delta < -20^\circ$ or $\delta > 85^\circ$ observed from APO have a design radius of $0.9^\circ$.  At LCO, due to vignetting, we typically impose a limit of 0.8$^\circ$ on survey targets, though exceptions to these rules may exist in special cases.

At the center of each SDSS plate is a hole that blocks targets being placed \citep[e.g.,][]{Owen_1994_sdssfibers}.  At APO, plates have a central post that obscures targets within 1.6$\arcmin$, while the use of a central acquisition camera at LCO (which also acts as a post) prevents targets within 5.5$\arcmin$ from the plate center.

The other large class of fields are those in which APOGEE targets are drilled and observed on plates driven by the MaNGA survey of resolved galaxies \citep{Bundy_2015_manga}.  Because of the MaNGA galaxy selection, these are located towards the Galactic caps, and the APOGEE target selection there is similar to APOGEE-led halo pointings (\S\ref{sec:main_sample}).  APOGEE does not control the positioning or observing prioritization of these fields.  For MaNGA observing details, see \citet{Law_2015_mangastrategy}.

\subsection{Field Locations}
\label{sec:field_locations}

As in APOGEE-1, fields can be divided into two rough categories: ``grid'' pointings, 
which form a semi-regular grid in Galactic longitude and latitude over a particular projected component of the MW,
and non-grid pointings, which are placed on particular objects of interest (such as stellar clusters).

The locations of the grid fields in APOGEE-2 are driven by a number of considerations.
We note that the separation between Northern and Southern pointings 
below is largely driven by accessibility and time constraints only; that is,
the majority of the total APOGEE-2 field plan, especially the grid pointings,
was designed as a coherent strategy to explore the MW, and {\it then} divided by hemisphere
and slightly modified by time constraints (the North has $\sim$2$\times$ the total amount of observing time as the South).  
The details of this chronological development of the plan are omitted here,
but when considering the motivation for fields that are classified as ``North'' or ``South'',
it may be helpful to keep this in mind.

For the Northern program, fields probing the Galactic disk were placed to fill in the APOGEE-1 coverage, or to revisit fields and complete the sample of bright red giants at those locations.
Fields in the halo with a faint magnitude limit, the so-called ``long'' 24-visit fields, are placed
on the location of shorter, 3-visit APOGEE-1 fields \citep{Zasowski_2013_apogeetargeting}, to extend the sample at those
locations with a larger number of distant stars.  
Shorter APOGEE-2N halo fields are placed to fill in the grid of short APOGEE-1 halo fields.
The few APOGEE-2N bulge fields are dedicated to a pilot study of red clump targets (\S\ref{sec:bulge_rc}).

For the Southern program, we placed fields in the disk and halo to mirror the Northern fields, both APOGEE-1 and -2, as closely as possible.  
Because of the greater accessibility of the Galactic bulge at LCO,
its coverage is significantly expanded than what was achievable from APO.
We placed fields in a grid pattern that included revisits to APOGEE-1 pointings
and new locations, with a mixture of 1-, 3-, and 6-visit depths.  Longer (deeper)
fields were placed on key regions of the bulge where larger, fainter samples are expected
to be most useful --- the bulge/halo interface at $|b| \sim 10^\circ$, 
and along the Galactic 
major and minor axes, for instance.

Non-grid fields in both hemispheres target stars in particular structures.
Fields placed on stellar clusters (\S\ref{sec:clusters} \& \ref{sec:young_clusters}), 
{\it Kepler} or {\it K2} pointings (\S\ref{sec:asteroseismic}--\ref{sec:koi}), 
dwarf galaxies (\S\ref{sec:sat_galaxies}), tidal streams (\S\ref{sec:tidal_streams}), 
or certain ancillary progam targets (\S\ref{sec:ancillary}) fall in this category.
The exact center of these fields is chosen to include the maximum number of stars
of interest, accounting for the FOV of the field (\S\ref{sec:field_properties}).

We note that as APOGEE-2 is still ongoing, additional fields may be added before the end of SDSS-IV.  For example, because of rapid survey progress enabled by better-than-average weather, program expansions are being considered for the remaining years of bright time at APO.  These expansions may include additional fields, or additional observations of existing fields on disk/halo substructures, stellar clusters, or calibration targets, among others.  Future publications and the online documentation will describe these changes in detail if they have a significant impact on the targeting strategy or anticipated yields for any part of the APOGEE-2 sample.

\subsection{Fields, Cohorts, Designs, Plates}
\label{sec:design_defs}

APOGEE-2 targeting employs the same ``cohort'' strategy adopted in APOGEE-1 to maximize
both targeting sample size and magnitude range \citep[][]{Zasowski_2013_apogeetargeting}.

In summary: a ``cohort'' is a group of stars that are observed together on the exact same visits to their field.
Thus, cohorts are selected and grouped based on the number of visits each star is expected to receive;
most commonly, this is done by magnitude (e.g., a 12-visit field's bright targets might only need three visits, but the faintest targets need all 12),
but there are exceptions.  These exceptions include bright stars being targeted for variability, which would be placed into the longest available cohort (e.g., \S\ref{sec:substellar}) rather than a shorter one with stars of comparable magnitude.
For this reason, cohorts are referred to as ``short'', ``medium'', and ``long'' --- defined by number of visits relative to that of other cohorts in the field --- rather than ``bright'' or ``faint''.  Which types of cohort are present, and how many visits they correspond to, varies by field.

A ``design'' is composed of one or more cohorts, and refers to the set of stars 
(including observational calibration targets: \S\ref{sec:calibration_observation})
observed together in a given visit.  
A ``plate'' is the physical piece of aluminum into which holes for all targets in a design 
are drilled and then plugged with optical fibers leading to the spectrograph.
A plate has only one design, but a design can be drilled onto multiple plates.\footnote{For example, if the same set of stars are intended to be observed at a very different hour angle, which requires slightly different positions to be drilled in the plate for the fibers.}
More details can be found in the Glossary in Appendix~\ref{sec:glossary}, and \S2.1 and Figure~1 of \citet{Zasowski_2013_apogeetargeting}.

\subsection{Design Priorities}

The order in which fields are started and completed over the course of the survey 
depends on a number of factors.  
For a given night of observing, plates are selected
by SDSS scheduling software that takes into account both the plate's observability at 
various times of the night and the priority of the plate's design.
For example, 
if two plates have identical survey priority, but one is observable for one hour that night and the other for three hours, the first one will be chosen for that slot, because the second can be more easily observed later.

The ``priority'' of a design, as used in this section, is set by the survey.
APOGEE-2 has three primary categories of scientific priority for designs: 
``core'', ``goal'', and ``ancillary''.
The core designs are those that address APOGEE-2's primary science goals:
the bulge, disk, and halo grid pointings (\S\ref{sec:main_sample});
stellar clusters (\S\ref{sec:clusters}); a complete astroseismic sample in the
original {\it Kepler} field (\S\ref{sec:asteroseismic}); 
satellite galaxies (\S\ref{sec:sat_galaxies});
halo stream candidates (\S\ref{sec:tidal_streams});
and red clump and RRL stars in the bulge (\S\ref{sec:rrl} and \S\ref{sec:bulge_rc}).
``Goal'' designs are those that use APOGEE-2's unique capabilities to address
related questions: {\it Kepler} Objects of Interest (KOIs; \S\ref{sec:koi}),
young star-forming clusters (\S\ref{sec:young_clusters}), 
{\it K2} asteroseismic targets (\S\ref{sec:asteroseismic}), 
and substellar companions (\S\ref{sec:substellar}).
There are two other target classes labeled ``goal'' 
(M dwarfs and eclipsing binaries, \S\ref{sec:m_dwarfs} and \S\ref{sec:ebs}, respectively), 
but these targets are sparsely distributed on the sky and appear on designs driven
by other factors.
Finally, ``ancillary'' designs are those dominated by ancillary targets (\S\ref{sec:ancillary}); small numbers of ancillary targets may also appear on core and goal designs.

These design classifications set a relative priority level in the 
observation scheduling software, but in practice, observability constraints
(including weather) play a greater role in deciding whether a design will be
observed or not on any given night.  
The order in which designs are selected and drilled onto plates, 
and thus available to be observed, is set
by the APOGEE-2 team's desire for early and ongoing diverse science output.
Thus some fields were designed in their entirety early in the survey, and
others are gradually designed and observed over the course of several years,
in order to both observe a wide range of target types and build large statistical samples
within the observing limits of the survey.

The array of fields and targets (and their various stages of completion) that are
available in a given data release is a result of all of these factors.

\section{Targeting Flags}
\label{sec:targeting_flags}

Understanding how representative APOGEE-2's spectroscopic target sample is 
of the stellar populations in the survey footprint requires knowing 
how that sample was selected from the underlying population.
APOGEE-2 uses three 32-bit integers (``targeting flags'', composed of ``targeting bits'')
to encode the criteria that are used to select targets.
Every target in a given design is assigned one of {\it each} of these integers,
whose bits indicate which criteria were applied to select that particular target 
for that particular design (e.g., color limit, calibration target, ancillary program).  
Thus each target may have up to 96 bits of targeting information associated with it (currently, not all bits are in use).
These targeting flags are design-specific, 
because targets can be selected for different reasons in different designs covering
the same spatial location.  
In practice, this situation is rare, but we note the possibility for completeness.

As in APOGEE-1 and the online documentation, throughout this paper we will use
the shorthand notation ``${\rm APOGEE2\_TARGET}X={\rm A}$'' to indicate that the targeting flag
APOGEE2\_TARGET{\it X} has targeting bit A set.  Technically, this bit is set by adding
$2^{\rm A}$ to the targeting flag.  Frequently, targets will have multiple bits set,
and their final targeting flags are sums of these bits:
\begin{equation}
{\rm APOGEE2\_TARGET}X = \sum_{i=0}^{N}{2^{{\rm bit}(i)}}.
\end{equation}
A list of APOGEE-2's current targeting bits is given in Table~\ref{tab:targeting_bits}.  Targets observed during APOGEE-1 use the APOGEE\_TARGET{\it X} flags (note the difference between APOGEE\_$\ast$ and APOGEE2\_$\ast$); see the SDSS DR12 documentation and Table~1 of \citet{Zasowski_2013_apogeetargeting} for these definitions.

For example, a star that is chosen as a calibration cluster member (\S\ref{sec:clusters}) that was also
observed in APOGEE-1 (\S\ref{sec:calibration_surveys}) would have ${\rm APOGEE2\_TARGET2=10}$ and ${\rm APOGEE2\_TARGET2=6}$ set;
if this star also happens to have been simultaneously selected as a member of the random
red giant sample (\S\ref{sec:main_sample}), it will have the associated dereddening method and 
cohort color and magnitude limit bits set as well.  See the DR14 SDSS bitmask documentation\footnote{\url{http://www.sdss.org/dr14/algorithms/bitmasks/}} and APOGEE-2 targeting documentation\footnote{\url{http://www.sdss.org/dr14/irspec/targets/}} for examples of targeting bit usage.

Finally, we note that these are {\it targeting} flags --- 
they indicate why a particular object
was selected for spectroscopic observation.  They do not indicate the actual nature of
the object or include any information learned from those observations.  
For example, stars targeted as possible members of open clusters may turn out not be actual members,
and stars not targeted (and flagged) as possible members may in fact be members.
The targeting flags should be used to reconstruct the survey selection functions
and identify targets associated with certain programs (such as ancillary projects),
{\it not} to identify a comprehensive list of particular types of objects.

\begin{table*}[!ht] \tablewidth{0 pt}
\caption{APOGEE-2 Targeting Bits}
\tabletypesize{\footnotesize}
\begin{center}
\begin{tabular}{llllll}
\multicolumn{2}{c}{APOGEE2\_TARGET1} & \multicolumn{2}{c}{APOGEE2\_TARGET2} & \multicolumn{2}{c}{APOGEE2\_TARGET3} \\
{\it Bit} & {\it Criterion} & {\it Bit} & {\it Criterion} & {\it Bit} & {\it Criterion} \\
0 & Single $(J-K_s)_0 > 0.5$ bin & 0 & --- & 0 & KOI target \\
1 & ``Blue'' $0.5 < (J-K_s)_0 < 0.8$ bin & 1 & --- & 1 & Eclipsing binary \\
2 & ``Red'' $(J-K_s)_0 > 0.8$ bin & 2 & Abundance/parameters standard & 2 & KOI control target \\
3 & Dereddened with RJCE/IRAC & 3 & RV standard & 3 & M dwarf\\
4 & Dereddened with RJCE/WISE & 4 & Sky fiber & 4 & Substellar companion search target \\
5 & Dereddened with SFD $E(B-V)$ & 5 & External survey calibration & 5 & Young cluster target \\
6 & No dereddening & 6 & Internal survey calibration (APOGEE-1+2) & 6 & --- \\
7 & Washington+DDO51 giant & 7 & --- & 7 & --- \\
8 & Washington+DDO51 dwarf & 8 & --- & 8 & Ancillary target \\
9 & Probable (open) cluster member & 9 & Telluric calibrator & 9 & --- \\
10 & --- & 10 & Calibration cluster member & 10 & -- {\it QSOs} \\
11 & Short cohort (1--3 visits) & 11 & --- & 11 & -- {\it Cepheids} \\
12 & Medium cohort (3--6 visits) & 12 & --- & 12 & -- {\it The Distant Disk} \\
13 & Long cohort (12--24 visits) & 13 & Literature calibration & 13 & -- {\it Emission Line Stars}\\
14 & Random sample member & 14 & Gaia-ESO overlap & 14 & -- {\it Moving Groups} \\
15 & MaNGA-led design & 15 & ARGOS overlap & 15 & -- {\it NGC 6791 Populations} \\
16 & Single $(J-K_s)_0 > 0.3$ bin & 16 & {\it Gaia} overlap & 16 & -- {\it Cannon Calibrators} \\
17 & No Washington+DDO51 classification & 17 & GALAH overlap & 17 & -- {\it Faint APOKASC Giants}\\
18 & Confirmed tidal stream member & 18 & RAVE overlap & 18 & -- {\it W3-4-5 Star Forming Regions}\\
19 & Potential tidal stream member & 19 & APOGEE-2S commissioning target & 19 & -- {\it Massive Evolved Stars} \\
20 & Confirmed dSph member (non Sgr) & 20 & --- & 20 & -- {\it Extinction Law} \\
21 & Potential dSph member (non Sgr) & 21 & --- & 21 & -- {\it Kepler M Dwarfs} \\
22 & Confirmed Mag Cloud member & 22 & 1-m target & 22 & -- {\it AGB Stars} \\
23 & Potential Mag Cloud member & 23 & Modified bright limit cohort ($H<10$) & 23 & -- {\it M33 Clusters} \\
24 & RR Lyra star & 24 & Carnegie (CIS) program target & 24 & -- {\it Ultracool Dwarfs} \\
25 & Potential bulge RC star & 25 & Chilean (CNTAC) community target & 25 & -- {\it SEGUE Giants} \\
26 & Sgr dSph member & 26 & Proprietary program target & 26 & -- {\it Cepheids} \\
27 & APOKASC ``giant'' sample & 27 & --- & 27 & -- {\it Kapteyn Field SA57} \\
28 & APOKASC ``dwarf'' sample & 28 & --- & 28 & -- {\it K2 M Dwarfs} \\
29 & ``Faint'' target & 29 & --- & 29 & -- {\it RV Variables} \\
30 & APOKASC sample & 30 & --- & 30 & -- {\it M31 Disk} \\
\hline \\
\end{tabular}
\end{center}
\label{tab:targeting_bits}
\end{table*}

\section{Science Sample Target Selection} 

\subsection{Main Red Giant Sample}
\label{sec:main_sample}

The targeting strategy for the primary red giant sample is very similar to that in APOGEE-1. 
We begin with the 2 Micron All Sky Survey Point Source Catalog \citep[2MASS PSC;][]{Skrutskie_06_2mass} and 
add mid-IR photometry from the {\it Spitzer}-IRAC GLIMPSE \citep{Benjamin_05_glimpse,Churchwell_09_glimpses} and AllWISE \citep{Wright_10_WISE,Cutri_2013_allwise} catalogs.
After applying data quality limits to ensure reliable photometry\footnote{A note about crowding: The exclusion of stars with 2MASS neighbors within 6$^{\prime\prime}$ (at least 3$\times$ the radius of the APOGEE-2 fibers) guarantees that bright neighbors are absent. We have assessed the impact of unresolved faint neighbors by examining APOGEE-1 spectra and ASPCAP results for stars around bulge globular clusters (the most crowded environment APOGEE-2 targets) with deeper VVV PSF photometry.  We see no significant difference in the spectra or the ASPCAP results for stars with faint VVV neighbors and those without.  This environment is also a worst-case scenario because the 2MASS faint limit is brighter here than elsewhere in the catalog, so we conclude that unresolved background light is not a dominant source of uncertainty.} (Table~\ref{tab:data_quality_reqs}), 
we use the RJCE method to calculate dereddened $(J-K_s)_0$ colors \citep{Majewski_2011_RJCE}.  
In some cases, particularly in the low-extinction halo fields, integrated $E(B-V)$ values from the \citet[][hereafter SFD]{Schlegel_1998_dustmap} maps are used for dereddening instead of the RJCE values.
The method used for each target is stored in that object's targeting flags (\S\ref{sec:targeting_flags}).

\begin{table*}[hpbt]
  \begin{center}
  \caption{Main Red Giant Sample Data Quality Requirements}
  \begin{tabular}{l c}
Parameter & Requirement \\
  \hline
2MASS total photometric uncertainty for $J$, $H$, and $K_s$ & $\le$0.1 \\
2MASS quality flag for $J$, $H$, and $K_s$ & $=$``A'' or ``B'' \\
Distance to nearest 2MASS PSC source & $\ge$6$^{\prime\prime}$ \\
2MASS confusion flag for $J$, $H$, and $K_s$ & $=$``0'' \\
2MASS galaxy contamination flag & $=$``0'' \\
2MASS read flag & $=$``1'' or ``2'' \\
2MASS extkey ID & $=${\it Null} \\
Photometric uncertainty for IRAC [4.5$\mu$m] & $\le$0.1 \\
Photometric uncertainty for WISE [4.6$\mu$m] & $\le$0.1 \\
{\it chi} for $M$, $T_2$, and {\it DDO51} data & $<$3 \\
{\it $|$sharp$|$} for $M$, $T_2$, and {\it DDO51} data & $<$1 \\
\hline
  \end{tabular}
  \label{tab:data_quality_reqs}
  \end{center}
\end{table*}

Stellar cohorts in the disk, bulge, and some halo fields are selected from 
candidate pools defined by ranges of $(J-K_s)_0$ color and $H$ magnitude.
Bulge and halo field cohorts use a single color limit of $(J-K_s)_0 \ge 0.5$ and $(J-K_s)_0 \ge  0.3$, respectively.
MaNGA-led fields, which lie towards the Galactic halo, are treated like halo fields in this respect.

\begin{figure*}[!hptb]
\begin{center}
\includegraphics[angle=0,trim=0in 0in 0in 0in, clip, width=0.9\textwidth]{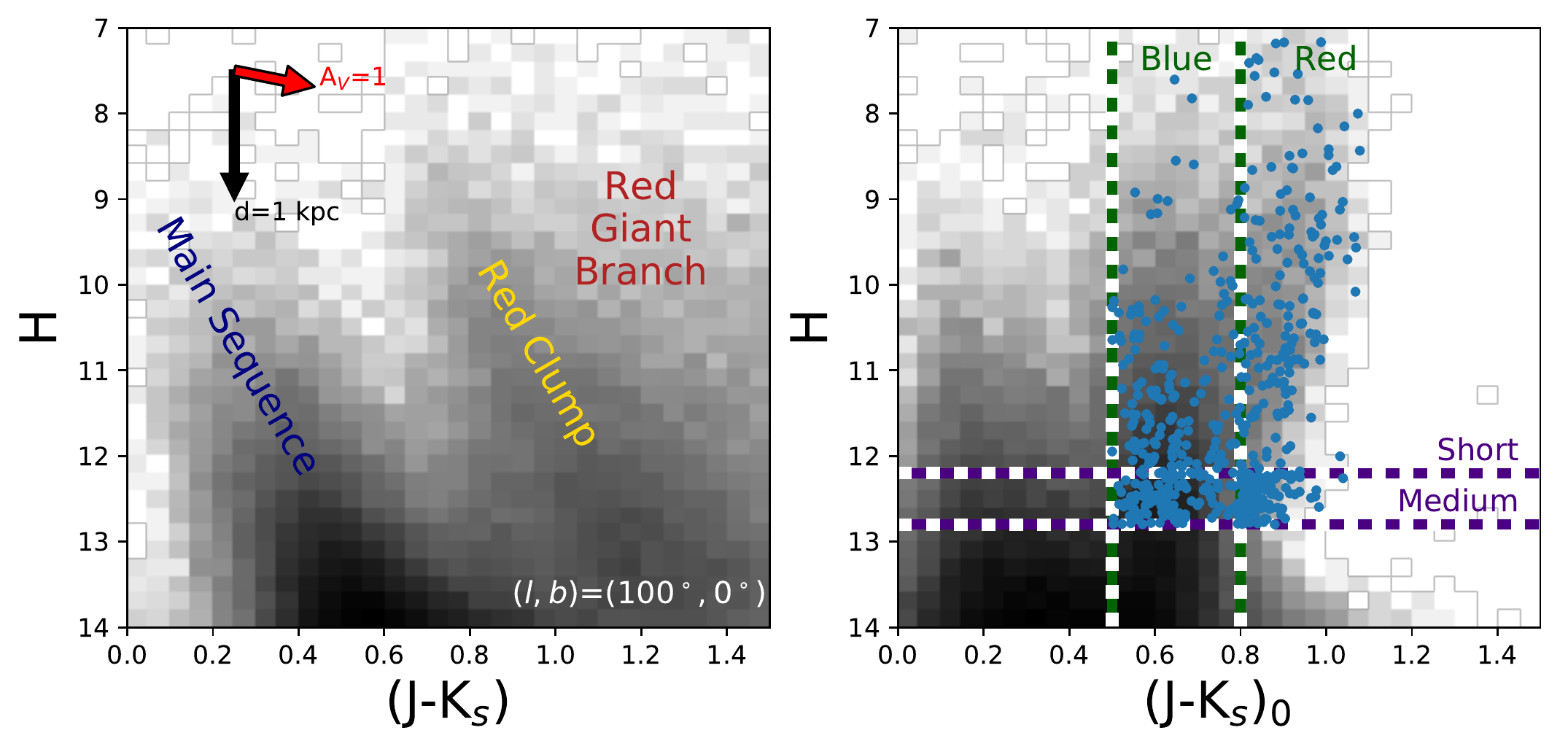} 
\caption{RJCE dereddening and target selection in the 6-visit disk field {\it 100+00}.  The left panel shows the observed $(J-K_s, H)$ CMD, with the primary populations labeled.  The right panel contains the same stars in the dereddened $([J-K_s]_0,H)$ CMD, overplotted with this field's targets (blue points).  As described in \S\ref{sec:main_sample}, targets are drawn from multiple bins in $H$ magnitude and dereddened $(J-K_s)_0$ color. 
}
\label{fig:diskcmd}
\end{center}
\end{figure*}

In a departure from APOGEE-1, the APOGEE-2 disk fields
utilize a two-color-bin scheme, with $N_{\rm blue}$ stars drawn from $0.5 \le (J-K_s)_0 \le 0.8$, and
$N_{\rm red}$ stars drawn from $(J-K_s)_0 \ge 0.8$.  This scheme is designed to increase the fraction of distant red giant stars.
The ratio between $N_{\rm blue}$ and $N_{\rm red}$ is set by the disk field's longitude: 
if the central longitude is $<120^\circ$ or $>240^\circ$, then half of that cohort's fibers are drawn from each bin (i.e., $N_{\rm blue} = N_{\rm red}$).
For outer disk fields with $120^\circ \le l \le 240^\circ$, the blue bin contains 25\% of the cohort fibers, and the red bin contains 75\%.  Figure~\ref{fig:diskcmd} illustrates these selection bins for one of the disk fields.
If there aren't enough stars available in a color bin, the ``extra'' fibers are assigned targets drawn from the other bin.

The faint magnitude limit of a cohort is set by the anticipated number of visits to that cohort, 
such that the faintest stars achieve the target summed S/N of 100 per pixel (see Table~\ref{tab:cohort_mag_limits}).
For Northern cohorts, these faint limits are identical to APOGEE-1.
The bright limit is set by the faint limit of shorter cohorts in the design, or by $H>7$ (the approximate saturation limit for a single visit) for the shortest cohort.  For some disk fields, a fainter bright limit of $H>10$ is adopted to avoid very nearby stars; targets in these designs have targeting bit APOGEE2\_TARGET2=23 set.
All of the designs in MaNGA-led fields were anticipated to be 3-visit cohorts, 
but with a slightly brighter faint limit of $H<11.5$ to account for
flux loss due to MaNGA's arcsec-sized spatial dither pattern (Table~\ref{tab:cohort_mag_limits}).

\begin{table}[hpbt]
  \begin{center}
  \caption{Typical $H$ Magnitude Limits of Primary APOGEE Cohorts}
  \begin{tabular}{c c}
$N_{\rm visits}$ & $H_{\rm min}-H_{\rm max}$ \\
  \hline
1 & 7.0--11.0 \\
3 & 7.0--12.2\footnote{7.0--11.5 in MaNGA-led fields} \\
6 & 7.0--12.8 or 12.2--12.8 \\
12 & 12.8--13.3 \\
24 & 13.3--13.8 \\
  \end{tabular}
  \label{tab:cohort_mag_limits}
  \end{center}
\end{table}

Furthermore, a number of the Northern halo fields have additional target
selection and prioritization based on stellar colors in the 
Washington $M$ \& $T_2$ and {\it DDO51} filters (hereafter ``W+D photometry'').
This technique takes advantage of the fact that 
because of the gravity sensitivity of the {\it DDO51} filter,
dwarf and giant stars form distinct loci in the $(M-T_2)$ vs.\ $(M-DDO51)$
color-color plane over a wide range of stellar temperature \citep[e.g.,][]{Majewski_00_W+Dtech}.
The application of this technique to select and prioritize giant stars 
in APOGEE targeting is described in \S4.2 of \citet{Zasowski_2013_apogeetargeting},
and APOGEE-2 uses the same classification method and subsequent observing priorities.  
That is, when selecting stars for a cohort with given color and magnitude limits,
stars classified as ``giants'' based on their W+D photometry are chosen first;
if any fibers allocated to that cohort remain, stars without W+D photometric classifications
are chosen next, followed by those classified as ``dwarfs.''

Proper motions for the APOGEE-2 main sample stars are drawn from the URAT1 catalog \citep{Zacharias_2015_urat1} and used to correct the target positions to the proper epoch before drilling the plates.  No proper motion data are used in the selection or prioritization of main sample targets.  APOGEE-2 observations later in the survey may use other proper motion catalogs, as improved ones become available \citep[e.g.,  {\it Gaia} and UCAC5;][]{GaiaCollab_2016_gaiaDR1,Zacharias_2017_ucac5}.

We emphasize that the color, magnitude, and W+D photometric criteria described here apply to stars 
selected as part of the primary red giant sample only.
This component comprised $\sim$2/3 of the total APOGEE-1 sample, and
we anticipate a similar fraction of the total APOGEE-2 sample.
The exact final proportion of main-sample red giant stars will depend on, e.g., the presence of additional ancillary programs, re-allocation of bright time made available by rapid observing progress, or other survey improvements.
Users of the main sample data should always check the documentation for the relevant data release for any updates to these criteria.
The subsections below describe the selection procedures for other
components of the APOGEE-2 survey.

\subsection{Open and Globular Clusters}
\label{sec:clusters}

Stellar clusters are valuable targets for chemical or dynamical surveys of the MW.  They provide a large number of stars with nearly identical ages, distances, and velocities, which can thus be measured more accurately and precisely than for isolated field stars.  Despite this, globular clusters are known to generally host multiple stellar populations \citep[e.g.,][]{mp_milone}, with well-characterized patterns in certain elemental abundances \citep[see][for a review]{Gratton_2012_multipleGCpops}.  In contrast, open clusters are generally considered to represent single populations with internally consistent abundances \citep[e.g.,][but see also \citet{Liu_2016_inhomogeneityM67,Liu_2016_inhomogeneityHyades}]{oc_jobovy,Ness_2017_stellardoppelgangers}.  Together, clusters represent star formation histories in a range of mass, metallicity, and Galactic environment.

Globular clusters in particular have been extensively studied with both photometry and spectroscopy, and this wealth of dynamical and chemical literature provides valuable benchmarks to calibrate newly derived datasets onto existing scales.  In addition, because clusters have little internal spread in age and (generally) in [Fe/H], 
their RGBs are useful for calibrating the behavior of $T_{\rm eff}$ and $\log{g}$ at fixed abundance.

For all of these reasons, open and globular clusters are targeted by APOGEE-2 for both scientific analysis and calibration.  APOGEE-1 observed a benchmark set of the globular clusters, and many of the open clusters, accessible in the Northern Hemisphere \citep[see][for calibration details]{Meszaros_2013_aspcapcalib, Holtzman_2015_apogeedata}.  In APOGEE-2, we revisit some Northern globular clusters (including M5, PAL~5, M12, M15, and M71) to increase the number of observed members.  In addition, the circumpolar open cluster NGC~188 is periodically observed from the North to monitor any long-term changes in the survey data properties, and other systems (including the open cluster NGC~2243) are observed with both the Northern and Southern instruments for internal cross-calibration (\S\ref{sec:calibration_surveys}).  

\begin{figure}[!hptb]
\begin{center}
\includegraphics[angle=0,trim=0in 0in 0.5in 0.5in, clip, width=0.45\textwidth]{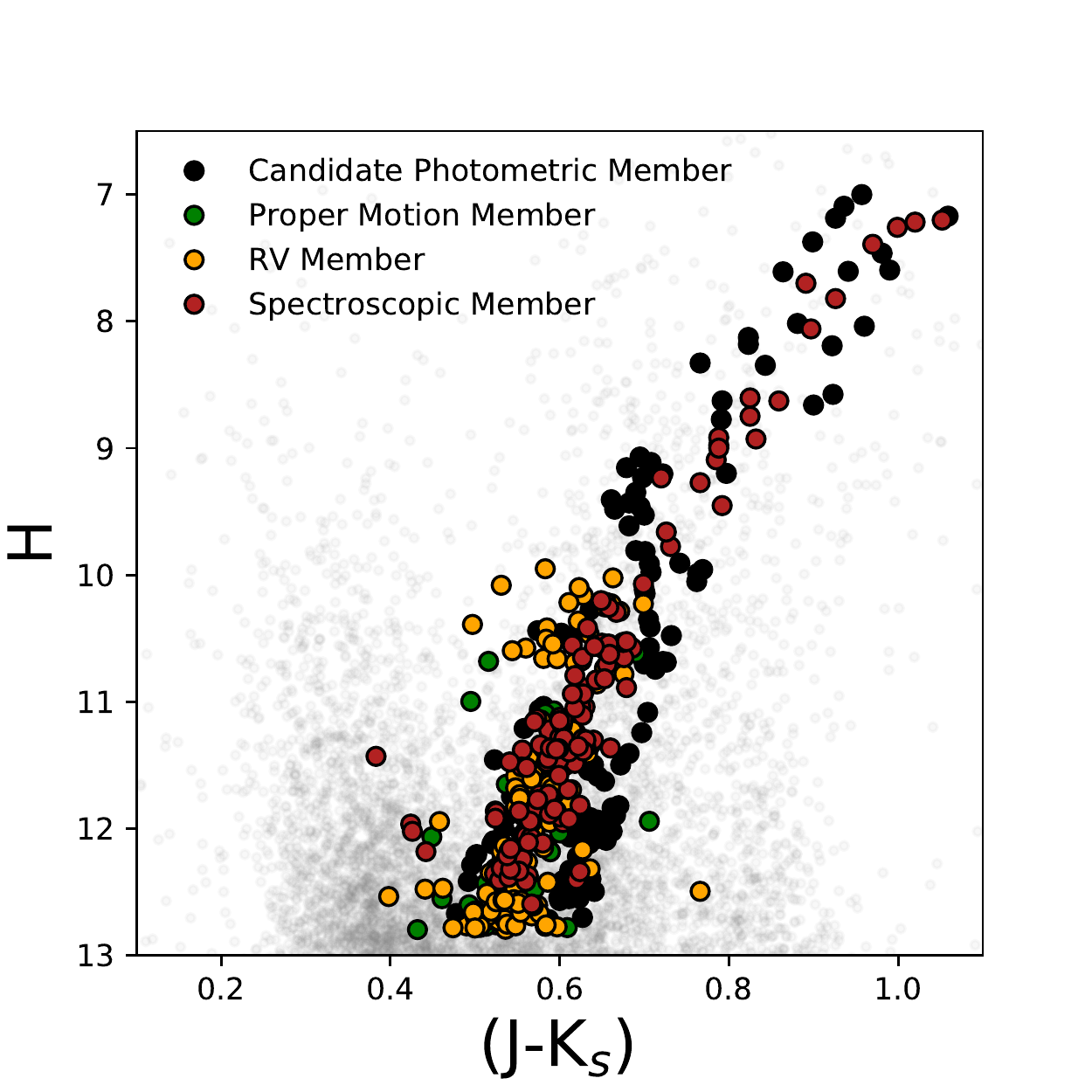} 
\caption{
CMD of targets in the globular cluster NGC~6752 and their priority classes, from stars with 2MASS colors consistent with the cluster locus (lowest priority, black circles), through known members with previous spectroscopic information (highest priority, red circles).  The faint gray points in the background are the rest of the 2MASS sources in the field that were not targeted as part of the cluster selection.
}
\label{fig:globclustercmd}
\end{center}
\end{figure}

APOGEE-2S is targeting a large number of globular and open clusters that are inaccessible to APOGEE-1 and APOGEE-2N.  Dedicated survey fields are placed on well-studied globular clusters with substantial pre-existing literature parameters, listed in Table~\ref{tab:clusters}, and 
another $\sim$10 globular clusters located towards the inner Galaxy fall at least partially within planned APOGEE-2S disk or bulge survey fields.   Targets in these clusters are selected and prioritized using a combination of 
pre-existing information and observing constraints imposed by APOGEE-2S's magnitude and fiber collision limits.  As in APOGEE-1, targeted members are selected according to the following priorities:
\begin{enumerate} \itemsep -2pt
\item presence of stellar atmospheric parameters and/or abundances derived from high-resolution data,
\item RV membership,
\item proper motion membership, and
\item location in the color-magnitude diagram (CMD), guided by the CMD locus of known members according to the previous criteria.
\end{enumerate}
Figure~\ref{fig:globclustercmd} shows the targets chosen with these selection categories for the globular cluster NGC~6752.

These targets are then sorted into cohorts (\S\ref{sec:design_defs}) to maximize the sample size by eliminating fiber colllisions.  Additional fibers are filled with cluster members (according to the above criteria) fainter than the nominal magnitude limit for the field or with field stars selected as part of the main survey's red giant sample (\S\ref{sec:main_sample}).  Stars selected as calibration cluster members (based on literature spectroscopic parameters and/or proper motion or RV membership probabilities) have the targeting bit APOGEE2\_TARGET2=10 set, and stars selected because they have high quality literature parameters or abundances have bit APOGEE2\_TARGET2=2 set. Note that these two flags are not mutually exclusive.  

Open clusters without significant literature of individual members are generally targeted using the selection algorithm described in \citet{Frinchaboy_2013_apogeeOCs}.  In summary, this is a spatial and photometric selection that uses information from the stellar sky positions, line-of-sight reddening, and proximity to a cluster isochrone derived from previous work (if known) to identify the likeliest cluster members.
Stars selected as potential open cluster members have the targeting bit APOGEE2\_TARGET1=9 set.
Analysis of APOGEE-1's open clusters began with \citet{Frinchaboy_2013_apogeeOCs} and \citet{Cunha_2015_occam}, and APOGEE-2 South is extending this large, homogeneous sample with $\gtrsim$100 clusters in the rich star formation regions of the southern Galactic disk.

\begin{table*}[hpbt]
  \begin{center}
  \caption{Observed and Anticipated Calibration Clusters}
  \begin{tabular}{l l c c c}
Cluster Name & NGC ID & [Fe/H]\footnote{\cite{carretta_feh}} & Distance\footnote{\cite{Harris_1996_MWGCs,Harris_2010_newMWGCs}} (kpc) & APOGEE-2 Field\footnote{All APOGEE-2S fields, with the exception of {\it M5PAL5}} \\
  \hline
  \\
M5 & NGC 5904 & $-1.33 \pm 0.02$  & 7.5  & {\it M5PAL5} \\
47Tuc & NGC 104 & $-0.76 \pm 0.02$ & 4.5  & {\it 47TUC} \\
      & NGC 288 & $-1.32 \pm 0.02$ & 8.9  & {\it N288} \\
      & NGC 362 &  $-1.30 \pm 0.04$ & 8.6  & {\it N362} \\
      & NGC 1851 & $-1.18 \pm 0.08$ & 12.1 & {\it N1851} \\
M79   & NGC 1904 & $-1.58 \pm 0.02$ & 12.9 & {\it M79} \\
      & NGC 2808 & $-1.28 \pm 0.04$ & 9.6  & {\it N2808} \\
      & NGC 3201 & $-1.51 \pm 0.02$ & 4.9  & {\it N3201} \\
M68 & NGC 4590 & $-2.27 \pm 0.04$ & 10.3  & {\it M68} \\
$\omega$ Cen & NGC 5139 & $-1.64 \pm 0.09$& 5.2  & {\it OMEGACEN} \\
M4 & NGC 6121 & $-1.18 \pm 0.02$ & 2.2  & {\it M4} \\
M12 & NGC 6218 & $-1.33 \pm 0.02$ & 4.8 & {\it M12} \\
M10 & NGC 6254 &  $-1.57 \pm 0.02$ & 4.4 & {\it M10} \\
 & NGC 6388 & $-0.45 \pm 0.04$ & 9.9 & {\it N6388} \\
 & NGC 6397 & $-1.99 \pm 0.02$ & 2.3 & {\it N6397} \\
 & NGC 6441 & $-0.44 \pm 0.07$ & 11.6 & {\it N6441} \\
M22 & NGC 6656 & $-1.70 \pm 0.08$ & 3.2 & {\it M22} \\
 & NGC 6752 & $-1.55 \pm 0.01$ & 4.0 & {\it N6752} \\
M55 & NGC 6809 & $-1.93 \pm 0.02$ & 5.4 & {\it M55} \\
  \end{tabular}
  \label{tab:clusters}
  \end{center}
\end{table*}


\subsection{Asteroseismic Targets}
\label{sec:asteroseismic}

Precise, high-cadence time-series photometry and spectroscopy of stars reveals that the surfaces of even non-variable, seemingly inactive stars fluctuate in response to standing waves reverberating throughout the star's layers.  Just as in Earth-based seismology, these wave patterns are affected by the density of the layers they encounter.  Through the analysis of these ``solar-like oscillations'' in a star, the fundamental stellar parameters of mass and radius (thus surface gravity) can be determined with high precision; when combined with spectroscopic temperature and abundance measurements, the {\it age} of a typical red giant star can be determined with $\sim$30\% accuracy.  This combination of data also provides masses for stars of known $T_{\rm eff}$ and abundance, rotation-based ages for dwarf stars via gyrochronological relationships, and diagnostics of internal stellar structure and pulsation mechanisms.

Highly precise, asteroseismic measurements of fundamental stellar parameters are valuable calibrators for less precise spectroscopic measurements, and benchmarks for models of stellar interiors and stellar atmospheres.  Stars with both asteroseismic and spectroscopic data are very useful for addressing numerous Galactic astrophysical questions, including the chemical and dynamical evolution of stellar populations and the demographics of transiting exoplanet host stars. In APOGEE-1, targets from two space satellites with asteroseismic data were observed. \citet{Anders_2017_corotgee} describes the APOGEE observations of $\sim$600 red giants with asteroseismic data from the {\it CoRoT} satellite \citep{Baglin_2006_corot}.  The APOGEE-{\it Kepler} Asteroseismic Science Consortium sample \citep[APOKASC sample;][]{Pinsonneault_2014_apokasc} includes more than 6000 giant and 400 dwarf stars with measured asteroseismic properties from the {\it Kepler} satellite \citep{Borucki_2010_keplerintro}.  

The initial APOKASC sample came from {\it Kepler}'s original pointing in Cygnus.  The bulk of the stars were selected based on their brightness, their photometric temperature, and the detection of solar-like oscillations in their light curves.  \citep[For details of the rest of the Cygnus APOKASC targets, see \S8.3 in][]{Zasowski_2013_apogeetargeting}. 

The APOGEE-2 APOKASC program is expanding this selection to build a magnitude-limited sample of {\it Kepler} stars, which contains significant improvements over the earlier subset.  In APOGEE-1, potential targets meeting the selection criteria were prioritized in ways that bias towards certain kinds of stars (e.g., first ascent RGB stars were preferred over RC stars), and the selection criteria themselves eliminated interesting stars (e.g., RGB stars {\it without} solar-like oscillations).  In addition, the stellar parameter spece spanned by the initial APOKASC catalog is relatively sparsely sampled.  

These limitations were unavoidable, given APOKASC's available time in APOGEE-1, and APOGEE-2 is dedicating a large amount of observing time to overcome them.  In the original Cygnus field, the remaining $\sim$13,500 cool stars with $7 \leq H \leq 11$ not yet observed with APOGEE are targeted, regardless of evolutionary state or the presence of oscillations.  Giants are identified using $T_{\rm eff} < 5500$~K and $\log{g} < 3.5$, and dwarfs with $5000 \leq T_{\rm eff} \leq 6500$~K and $\log{g} > 3.5$; pre-observation temperature and gravity estimates come from the revised Kepler Input Catalog \citep{Huber_2014_revisedKIC} and the corrected temperature scale of \citet{Pinsonneault_2012_KICTeffrevision}.  These targets are distributed across the same 21 fields used in APOGEE-1 APOKASC, and are observed with a total of $\sim$50 single-visit designs.

In 2013, the second of the {\it Kepler} spacecraft's four reaction wheels failed, leaving the satellite unable to remain stably pointed at the Cygnus field.  The telescope was then repurposed for the {\it K2} mission, performing very similar observations at numerous pointings along the ecliptic plane in 75-day intervals \citep{Howell_2014_K2}.  Though stellar asteroseismic measurements from these data are noisier than their counterparts from the $\sim$4~year {\it Kepler}-Cygnus data, the {\it K2} sample is highly valuable for Galactic stellar and planetary studies because it includes stars spanning an enormous range of Galactic environment, from stellar clusters and the halo to the bulge.  

APOGEE-2 is targeting more than 10$^4$ giant stars in several of these {\it K2} pointings (called ``campaigns''), mostly from the K2 Galactic Archaeology Program's (GAP) sample of asteroseismic targets.\footnote{\url{ http://www.physics.usyd.edu.au/k2gap/}}  These stars are selected from the pool of GAP candidates based purely on the their magnitude, with additional non-GAP stars selected as oscillators based on {\it K2} data. The total sample is observed over at least 50 visits divided amongst the campaigns, all or nearly all of which are being observed from the North.

All APOKASC targets have targeting bit APOGEE2\_TARGET1=30 set, with giants and dwarfs being further identified with APOGEE2\_TARGET1=27 and 28, respectively, if applicable.


\subsection{Kepler Objects of Interest}
\label{sec:koi}

The {\it Kepler} satellite has identified thousands of transiting planets confirmed through follow-up spectroscopy or imaging.  Prior to confirmation, a star with 
transit signals potentially consistent with orbiting planets
is called a ``{\it Kepler} Object of Interest'' (KOI).  As a class, KOIs include genuine planet hosts along with eclipsing binary systems, brown dwarf hosts, strongly spotted stars, and other systems with light curves that can masquerade as transiting planet signatures.  

High-resolution, high-cadence spectroscopy can
distinguish many of the ``false positive'' cases from true planets
\citep{Fleming_2015_KOIs}.
While the APOGEE RV precision is generally insufficient to detect planets directly, the data can identify several of the most common classes of false positives, including eclipsing binaries with grazing orbits, tertiary companions diluting binary system eclipse depths, and very low mass stars or brown dwarfs, with radii (and thus transit depths) similar to those of gas giant exoplanets but with much larger masses.  The RV signal of these types of systems ranges from several hundred m~s$^{-1}$ to tens of km~s$^{-1}$.
Using the {\it Kepler} planet sample to improve our understanding of planetary system formation and stellar host demographics requires 1) using multi-epoch RV data to strictly constrain the false positive rate as a function of stellar type, and 2) measuring the stellar properties of both planet-hosting and non-planet-hosting populations with significant statistics (see also \S\ref{sec:substellar}). 

APOGEE-2 is making observations to address both of these requirements, using samples of confirmed planet hosts, KOIs, and non-hosts in the {\it Kepler} Cygnus footprint.  The primary goal is to homogeneously measure the binary fraction, chemical compositions, and rotational velocities for significant samples of both KOIs and non-KOIs.  The first two properties are thought to have an impact on planet frequency and habitability, and the third can be used to constrain stellar activity and tidal effects, which also affect planetary system properties.  Furthermore, the spectroscopic RVs and rotational velocities can help discriminate among different sources of false positives to provide a cleaner planet sample, and some of these sources (such as brown dwarfs) are scientifically interesting objects in their own right.

The APOGEE-2 KOI program contains  
$\sim$1000 KOIs and $\sim$200 non-hosts distributed across five APOGEE-2 fields, supplemented by $\sim$200 KOIs observed in APOGEE-1.  Planet hosts and KOIs were drawn from the NExScI archive\footnote{\url{http://exoplanetarchive.ipac.caltech.edu}, queried immediately prior to the design of each field: March 2014 ({\it K10\_079+12, K21\_071+10}), March 2016 ({\it K04\_083+13}), and February 2017 ({\it K06\_078+16, K07\_075+17}).} using a simple magnitude limit of $H<14$ to identify all CONFIRMED or CANDIDATE targets in the fields.  The non-host ``control sample'' was drawn from the {\it Kepler} Input Catalog \citep{Brown_2011_KIC}, using the same $H<14$ magnitude limit and selected to provide the same $T_{\rm eff}$--$\log{g}$ joint density distribution as in the host+KOI sample.  These control sample stars are used to fill fibers unused by the host+KOI sample.  

Each APOGEE-2 KOI field is observed over 18 epochs, with cadencing sufficient to characterize a wide range of orbits.  The host+KOI targets can be identified with the targeting bit APOGEE2\_TARGET3=0, and the control sample targets with APOGEE2\_TARGET3=2.


\subsection{Satellite Galaxies}
\label{sec:sat_galaxies}

The bulk of the APOGEE-2 programs are dedicated to measuring chemo-dynamical properties of stars within the MW.  However, placing these properties in the context of the MW's evolution, and of galaxy evolution in general, requires comparable measurements of other stars in the Local Group.  APOGEE-2 is targeting stars in \textcolor{black}{eight} Local Group satellite galaxies: the Large and Small Magellanic Clouds (LMC, SMC; \S\ref{sec:mag_clouds}) and six dwarf spheroidal galaxies (dSphs; \S\ref{sec:dsphs}).

\subsubsection{Magellanic Clouds}
\label{sec:mag_clouds}

The LMC and SMC are the two most massive satellite galaxies of the MW, and at distances of only $\sim$50~kpc and $\sim$60~kpc, respectively, they span angles on the sky large enough for APOGEE to obtain velocity and chemical information for thousands of individual stars.  By mapping the kinematics and abundances of these stars, as well as the interplay between stellar and gas kinematics, across the galaxies and the MW, we can explore the mass dependence of chemical evolution and feedback in the range of $10^8-10^{10}~M_\sun$.

APOGEE-2 is targeting $\sim$3500 stars in $\sim$17 fields spanning the LMC down to a magnitude limit of $H=14.6$, and $\sim$2000 stars in $\sim$9 fields in the SMC down to $H=14.9$; two of the SMC fields include the MW globular clusters 47~Tuc and NGC~362.  
The LMC fields extend from the center of the LMC out to $\sim$9.5$^\circ$ along the major axis and $\sim$6.5$^\circ$ on the minor axis; the SMC fields span up to $\sim$5$^\circ$ away from the SMC's center along both axes.  These fields cover approximately 1/3 of the sky area within those ranges. Their exact placement is driven by the desire for well-sampled radial and azimuthal coverage, as well as for a high local density of MC RGB stars and the presence of DES or SMASH photometry \citep[][]{DESCollab_2016_DESoverview,Nidver_2017_smash}, which can provide star formation histories for the spectroscopic sample.

\begin{figure}[!hptb]
\begin{center}
\includegraphics[angle=0,trim=0in 0in 0.5in 0.5in, clip, width=0.45\textwidth]{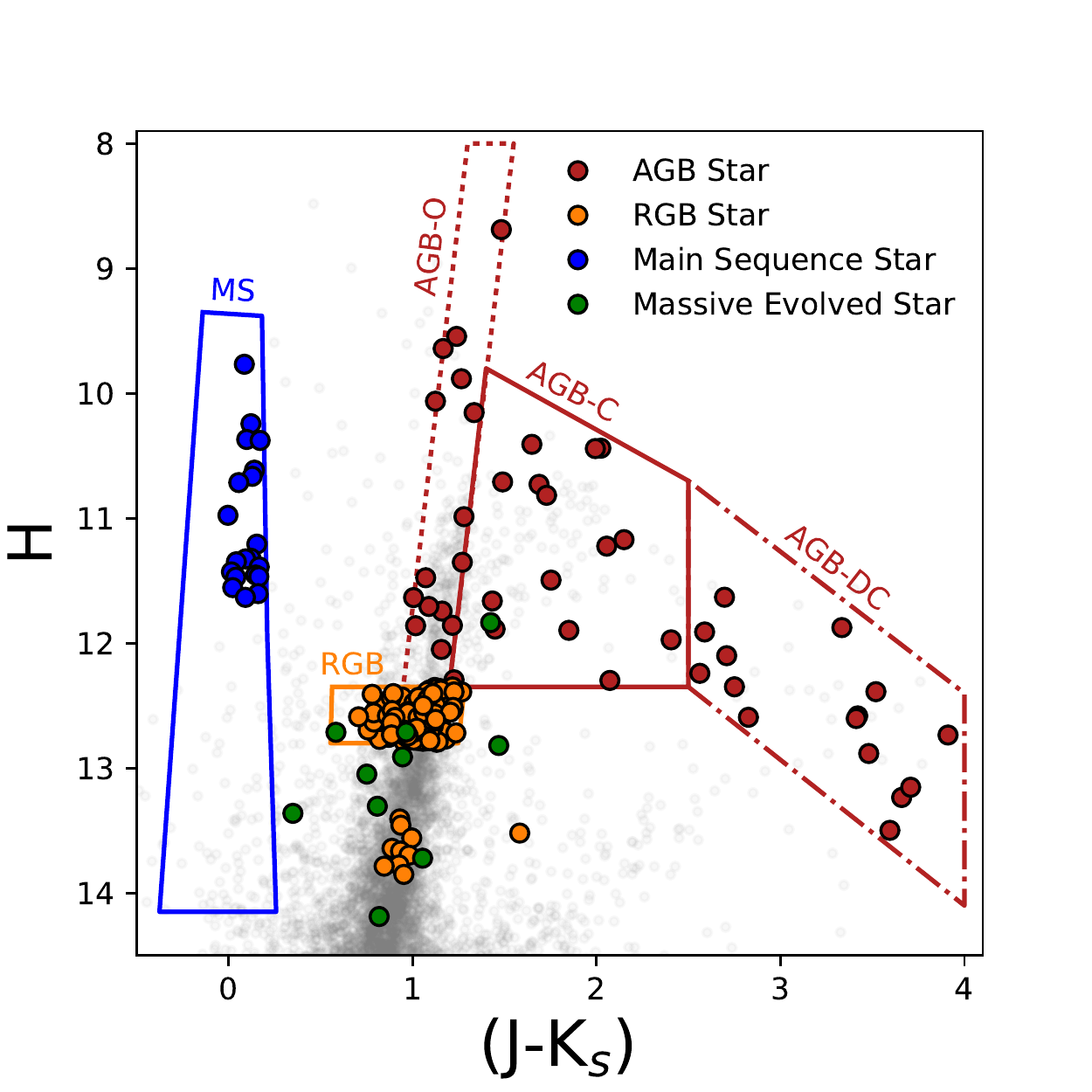} 
\caption{
CMD of targets and target classes in a sample LMC field ({\it LMC-9}).  Red points indicate AGB stars, comprising AGB-O, AGB-C, and dusty AGB-C (AGB-DC) stars appearing in three different color-magnitude bins.  Orange indicates RGB stars, which include both stars selected from the orange box and those with existing literature data.  Blue are hot main sequence stars, selected from the drawn box, and green are other massive evolved stars selected from the literature.  Gray points are other 2MASS sources in the field that were not targeted as LMC stars.
}
\label{fig:magcloudcmd}
\end{center}
\end{figure}

The MC targets comprise four primary types of stars: RGB stars, AGB stars, young massive stars, and other rare massive evolved stars (Figure~\ref{fig:magcloudcmd}).  RGB stars provide a dense sampling of the abundance gradients and chemical evolution of the MCs, including in the poorly understood SMC and outer regions of the LMC.  These outer pointings will also be used to look for tidal streams and other substructures in these low mass halos, which are predicted by hierarchical formation models; one such stream has already been detected \citep{Olsen_2011_LMCaccretion}, and APOGEE's numerous chemical abundances will be powerful tools for identifying others.

AGB stars, though poor probes of galactic chemical evolution because of the internal mixing that has modified their atmospheric abundances, are important tracers of {\it stellar} chemical evolution for the same reason.  Measurement of abundances that are expected to be altered by physical processes and AGB nucleosynthesis, compared to abundances of the less-evolved RGB stars, enable studies of the mass- and metallicity-dependence of processes such as third dredge-up, hot bottom burning, and the synthesis of elements such as N, C, Na, Al, and Mg \citep[e.g.,][]{ventura15_agb,ventura16_agb}.  At the other end of the stellar lifecycle, young hot stars in the MCs contain chemodynamical information from very recent and ongoing star formation, to compare with co-spatial gas measurements and with the more evolved stellar populations.  

APOGEE-2 targeting in the LMC and SMC uses extensive photometric and spectroscopic information to produce a large, clean set of diverse targets.  
Candidate stellar targets in LMC and SMC fields are divided into subclasses, including red giants, supergiants, massive young main sequence stars, AGB and post-AGB stars.  Subclasses are identified using ($J-K_{S},H$) color-magnitude selections as well as classifications made from {\it Spitzer}-IRAC color-color and color-magnitude diagrams, optimal for characterizing O- and C-rich AGB stars \citep{dellagli_agb1,dellagli_agb2}.  Candidate target lists in each LMC/SMC field have also been cleaned of foreground Milky Way dwarfs using color-color limits in W+D photometry.  Figure~\ref{fig:magcloudcmd} shows the target selection in an example LMC field.  We note that, as for young stars targeted elsewhere in the survey (e.g., \S\ref{sec:young_clusters}), the automated spectral fits produced by ASPCAP may not be reliable for all of these targets, and customized analysis may be required.

Cohorts (\S\ref{sec:design_defs}) are not used for LMC/SMC targeting given the faint magnitudes of the targets, although the fraction of stars populating each subclass is tailored from field to field.  This allows the LMC/SMC target selection process to accommodate variations in relative stellar density on the sky, ensuring that intrinsically rare objects are observed where possible while simultaneously guaranteeing a sizeable sample of RGB and AGB targets over a range of luminosities.

Stars targeted as confirmed MC members based on existing high resolution spectroscopy are flagged with the targeting bit APOGEE2\_TARGET1=22, and those selected as likely members using photometry are flagged with APOGEE2\_TARGET1=23.

\subsubsection{Dwarf Spheroidals}
\label{sec:dsphs}

One of the factors inhibiting the measurement of chemical abundances for large samples of more distant dwarf galaxies is the faintness of their stars, which requires long exposures for a useful high-resolution spectrum.  APOGEE-2's multiplexing capability over a large FOV enables the collection of relatively large samples at a rate competitive with that of instruments on larger telescopes,
with the additional benefit that RGB stars in these systems are brighter in the $H$-band than at optical wavelengths.

APOGEE-2 is targeting $\lesssim$200 stars in each of six dSphs: Ursa Minor, Draco, and Bo\"{o}tes~I with the APOGEE-2N, and Sculptor, Sextans, and Carina with APOGEE-2S.  Each of these fields is being observed for about 24 visits, with fainter stars (which dominate the sample) being assigned fibers on all 24 visits, while any brighter ones switched out after 6 or 12 visits (analogous to the main sample's cohort scheme; \S\ref{sec:design_defs}).  The primary goals of this dataset are to: (1) obtain larger chemical abundance samples in the target galaxies than have generally been possible in the past, (2) map any spatial population gradients, and (3) trace high-dimensional chemical patterns, especially in elements that are difficult to measure accurately at optical wavelengths for cool, metal-poor dSph stars, such as O and Si.  The data will also (4) enable the measurement of stellar binary fractions, determination of the orbits of identified binaries, and assessment of the impact of binary stars on dSph velocity dispersions and their inferred dark matter content.

\begin{figure}[!hptb]
\begin{center}
\includegraphics[angle=0,trim=0in 0in 0.5in 0.5in, clip, width=0.45\textwidth]{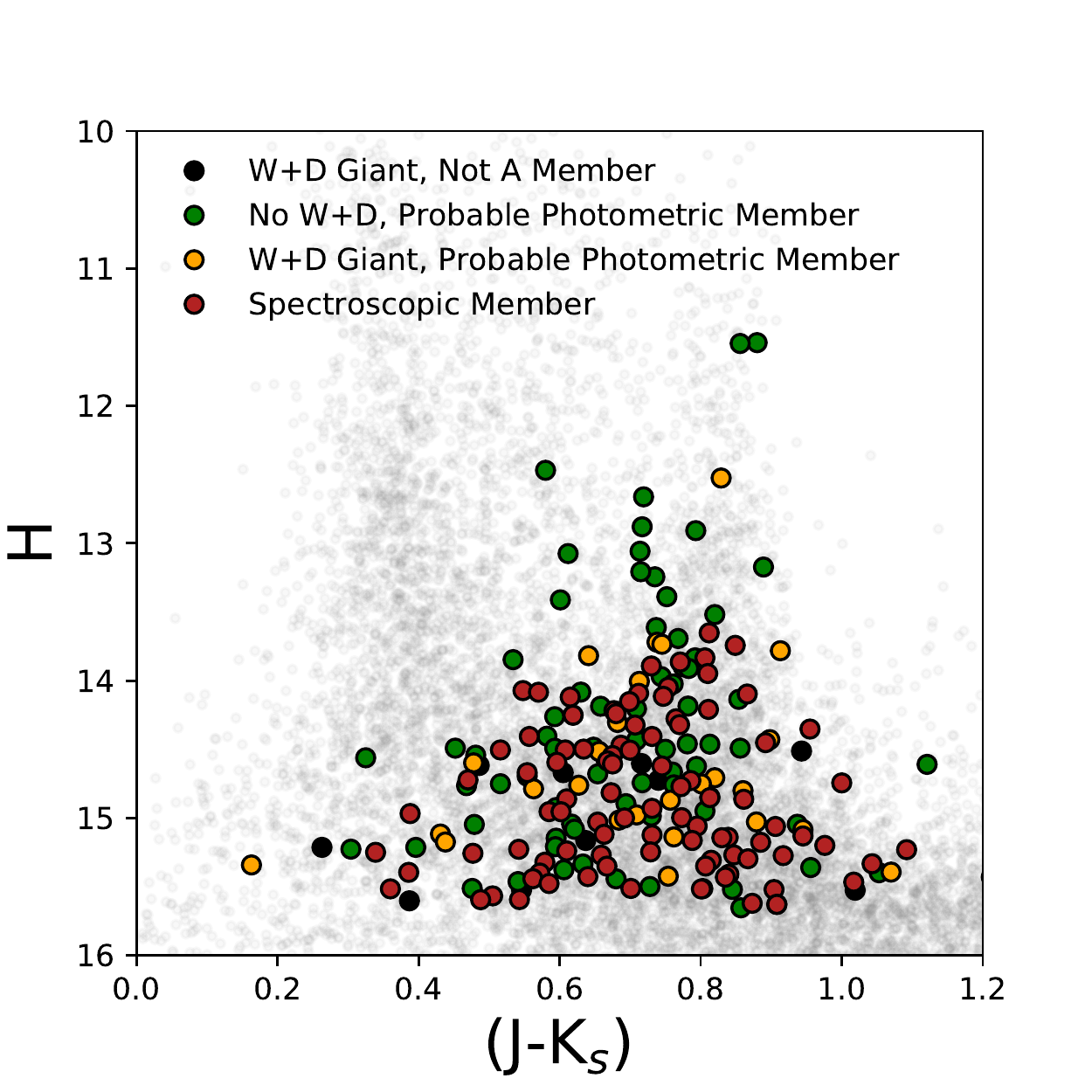} 
\caption{
CMD of targets in the Ursa Minor dwarf galaxy and their priority classes, from W+D giants with broadband colors {\it inconsistent} with galaxy membership (lowest priority, black circles), through known members with previous spectroscopic information (highest priority, red circles).  The faint gray points in the background are the rest of the 2MASS sources in the field that were not targeted as part of the dSph selection.
}
\label{fig:dsphcmd}
\end{center}
\end{figure}

Dwarf galaxy targets are selected with a range of spectroscopic and photometric criteria: W+D photometry, other broadband photometry where necessary (e.g., SDSS), and spectroscopic membership information based on radial velocities from the literature.  The highest priority targets are previously confirmed member stars.  Of next highest priority are stars classified as giants from their $M$, $T_2$, and {\it DDO51} colors (similar to the grid halo fields; \S\ref{sec:main_sample}) and with broadband colors that place them near the RGB of the dwarf galaxy.  For any portions of the field lacking W+D imaging, broadband colors alone are used.  
Remaining fibers are allocated to W+D photometric giants that are not consistent with the dwarf galaxy RGB or to stars selected under the general halo targeting criteria (\S\ref{sec:main_sample}). 
Figure~\ref{fig:dsphcmd} shows the targets chosen with these selection categories for the Ursa Minor dwarf galaxy (in the {\it URMINOR} field).
Further details of the targeting for specific galaxies will be described in future papers.

Stars selected as targets in these dwarf galaxies using 
spectroscopic membership information are flagged with the targeting bit APOGEE2\_TARGET1=20, and those from photometric criteria alone are flagged with APOGEE2\_TARGET1=21.  

\subsection{Tidal Stream Candidate Members}
\label{sec:tidal_streams}

Though it contains only a tiny fraction of the Galaxy's total number of stars, the Galactic halo contains a number of evolutionary fossil records: stellar streams that are the remnants of galactic mergers and dissolving stellar clusters.  APOGEE-2 is targeting a number of these structures to obtain the chemodynamical information necessary to determine their origin in the context of the halo's history.

The Triangulum-Andromeda (TriAnd) structure is a diffuse, $\sim$2000~deg$^2$ overdensity detected in MSTO stars in the foreground of M31 \citep{Majewski_2004_TriAnd,Rocha-Pinto_2004_TriAnd}.
TriAnd's nature as either a satellite remnant or disk substructure, and/or the superposition of multiple structures, remains unresolved \citep[e.g.,][]{Martin_2007_TriAnd,Chou_2011_TriAndChem,Xu_2015_MWDiskRings,Price-Whelan_2015_TriAnd}.  APOGEE-2 is targeting five locations where the standard halo selection criteria, without W+D photometry (\S\ref{sec:main_sample}), select the maximum number of TriAnd member candidates from \citet{Sheffield_2014_TriAnd} and \citet{Chou_2011_TriAndChem}.  These fields are called {\it TRIAND-1, -2, -3, -4}, and {\it 5}.  Both member and non-member stars with chemical analyses from \citet{Chou_2011_TriAndChem} are included for comparison.

Four additional streams are targeted using a variety of data to identify likely members: Pal-5 \citep{Odenkirchen_2001_Pal5tails,Odenkirchen_2002_Pal5spectra,Odenkirchen_2003_Pal5tails,Odenkirchen_2009_Pal5kinematics,Carlberg_2012_Pal5gaps}, GD-1 \citep{Koposov_2010_GD1}, Orphan \citep{Newberg_2010_orphanstream,Sesar_2013_orphanstreamRRL,Casey_2013_orphanstream,Casey_2014_orphanstream}, and the Sgr tail \citep{Bellazzini_2003_Sgrstream,Yanny_2009_Sgr,Carrell_2012_SgrRCstars,PilaDiez_2014_Sgrwraps}.  Each stream has 1--5 numbered fields placed along its length (e.g., {\it GD1-1}, {\it GD1-2}, etc).  The highest likelihood candidate members are identified as those stars (1) classified as ``giant'' using W+D photometry (\S\ref{sec:main_sample}), and (2) falling close to the stream's expected locus in a ([$J-K_s]_0$, $H$) CMD.  
Reddening values for these stars are drawn from the SFD $E(B-V)$ maps, and distance and metallicity information is taken from the references listed above.  The isochrones of appropriate metallicities are from the PAdova and TRieste Stellar Evolution Code \citep[PARSEC;][]{Bressan_2012_parsec}.

At lower priority than the W+D-classified giants matching their stream's distance-shifted isochrone in the 2MASS CMD, we targeted stars matching the isochrone but without a dwarf/giant classification, then giants farther from the isochrone, then unclassified stars farther from the isochrone.  Stars too faint to obtain sufficient S/N in their field, very red ($[J-K_s]_0>1.3$), and very blue ($[J-K_s]_0<0.2$) stars are eliminated, along with any stars having UCAC4 proper motions well outside the uncertainties of the stream's proper motion at that position (if known).

Stream targets have targeting bit APOGEE2\_TARGET1=19 set, along with applicable W+D photometric classification flags.


\subsection{Embedded Young Clusters}
\label{sec:young_clusters}

APOGEE-2 is targeting a number of deeply embedded young stellar clusters to characterize the earliest stages of the older populations dominating the rest of the sample.  High-precision ($<$1~km~s$^{-1}$) RVs and fundamental stellar parameters are difficult to measure in these extremely extinguished environments, so APOGEE-2's IR sensitivity and multiplexing capability are ideally suited to providing a systematic census of the dynamics and binary fractions of these clusters.  

The primary science drivers of this target class are to characterize the dynamical processes that drive the formation, evolution, and (usually) disruption of young stellar clusters, and to constrain the frequency and properties of close binaries in these systems.  These data will also be useful for refining the global properties of each cluster (e.g., age, distance, IMF), measuring the magnetic field strengths of pre-main sequence (PMS) stars \citep[e.g.,][]{JohnsKrull_2009_BFieldClassI}, constraining both average and variable accretion rates, and testing PMS evolutionary tracks.

\begin{figure}[!h]
\begin{center}
\includegraphics[angle=0,trim=0in 0in 0in 0.4in, clip, width=0.48\textwidth]{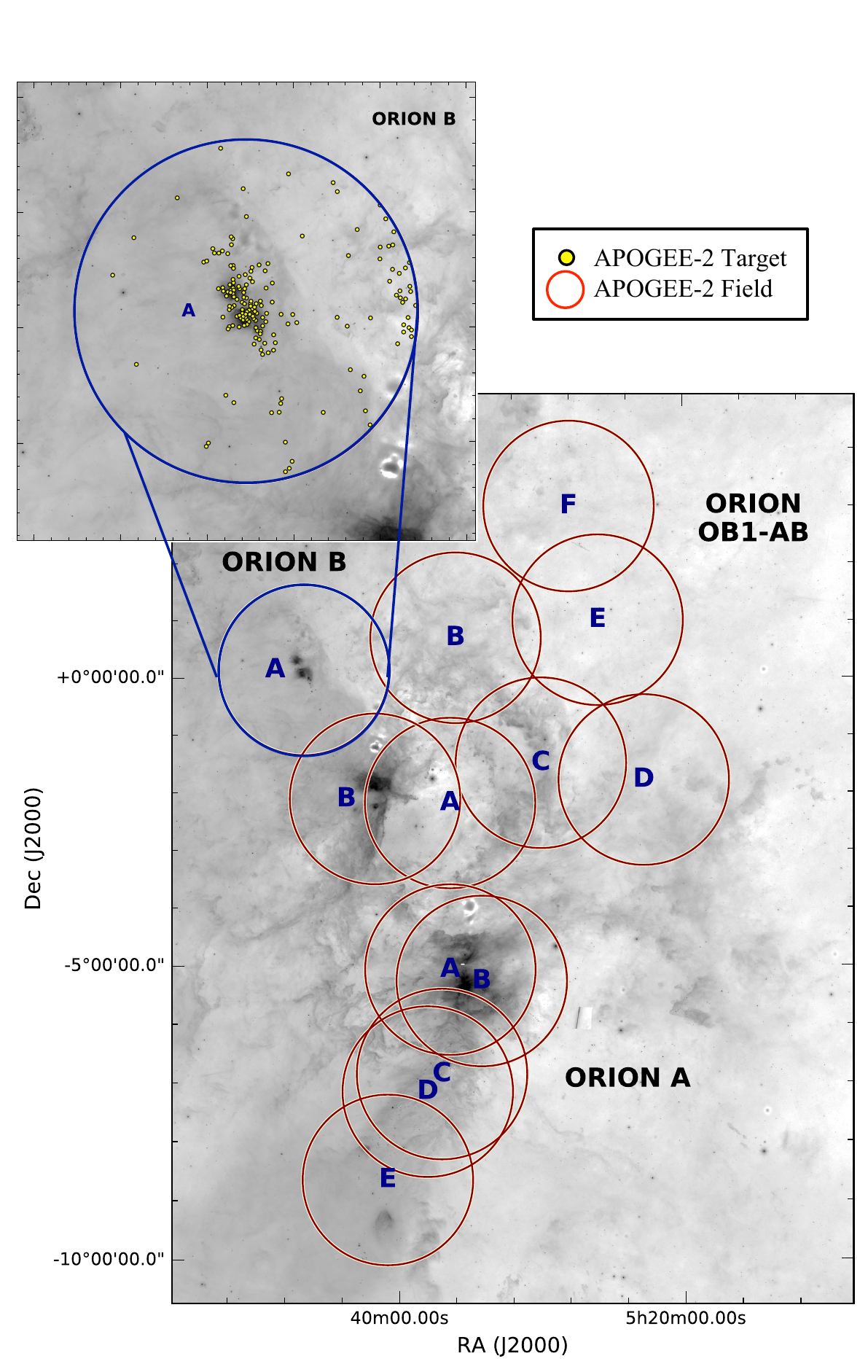} 
\caption{
Example of targeting in embedded young clusters, here for the extended Orion fields: Orion~A, Orion~B, and Orion~OB1-AB.  The background grayscale is {\it WISE} 12$\mu$m emission, and the red circles indicate the location of APOGEE-2 fields.  The letters in each field identify the field name --- e.g., {\it ORIONB-A} and {\it ORIONA-E}. The inset shows targets in the {\it ORIONB-A} field.
}
\label{fig:insync}
\end{center}
\end{figure}

To these ends, APOGEE-2 is targeting approximately 200--1000 sources in each of several embedded cluster complexes, which are listed in Table~\ref{tab:insync_fields}.  An example is shown in Figure~\ref{fig:insync}.  This target class is an extension of the young stellar cluster program from APOGEE-1 \citep[IN-SYNC;][]{Cottaar_2014_insync} and shares similar targeting procedures.  The need for multiple epochs to measure the apparent RV variability (``jitter'', e.g., due to spots and outflows) of young stars drives a magnitude limit of $H \leq 12.5$; brighter stars, which have more precise single-epochs RVs, are switched out after a smaller number of visits (similar to the main sample cohort scheme; \S\ref{sec:design_defs}).  The cluster targets are drawn from compiled catalogs of sources with optical/IR photometry consistent with the cluster locus, along with IR excesses, X-ray activity, Li abundance, H$\alpha$ excess, and variability consistent with that seen in young stars (J.~Cottle et al., submitted).  Targets are prioritized by brightness and by spatial distribution within each cluster to maximize the sampling of the velocity structure.  The finer details of targeting for specific clusters will be described in the associated papers.

Note that APOGEE's ASPCAP pipeline does not include PMS stellar models, so the automated synthetic spectral fits are not likely to be meaningful for most of these sources.  A customized analysis pipeline has been developed for these young objects, which includes PMS-specific considerations such as accretion signatures and flux from the circumstellar disk \citep{Cottaar_2014_insync,DaRio_2016_insyncOrionA}.  

Sources targeted as part of the young cluster program are flagged with APOGEE2\_TARGET3=5.  

\begin{table*}[hpbt] \tablewidth{0 pt}
  \begin{center}
  \caption{Embedded Clusters Targeted in APOGEE-2}
  \begin{tabular}{c c c c l}
Cluster\footnote{Additional clusters may be targeted; see the data release documentation for final details.} & Age [Myr] & Distance [pc] & Number of Targets\footnote{Approximate number, and anticipated counts for fields not yet observed as of the writing of this paper.} & APOGEE-2 Field(s) \\
  \hline \\
Orion~A & 1--3 & 388-428 & 500 & {\it ORIONA-A, -B, -C, -D, -E} \\ 
Orion~B & 1--3 & 388-428 & 1000 & {\it ORIONB-A, -B} \\ 
Orion~OB1 & 5--10 & 388 & 3300 & {\it ORIONOB1AB-A, -B, -C, -D, -E, -F} \\ 
$\lambda$~Ori & 3--5 & 450 & 1900 & {\it LAMBDAORI-A, -B, -C} \\ 
Pleiades & 115 & 135 & 450 & {\it PLEIADES-E, -W} \\ 
Taurus~L1495 & 1--4 & $\sim$130 & 70 & {\it TAU1495} \\
Taurus~L1521 & 1--4 & $\sim$130 & 30 & {\it TAU1521} \\
Taurus~L1527 & 1--4 & 130--160 & 40 & {\it TAU1527} \\
Taurus~L1536 & 1--4 & $\sim$130 & 40 & {\it TAU1536} \\
$\alpha$~Per & 85 & 172 & 200 & {\it ALPHAPER} \\
NGC~2264 & 1--6 & $\sim$800 & 400 & {\it NGC2264}
  \end{tabular}
  \label{tab:insync_fields}
  \end{center}
\end{table*}


\subsection{Substellar Companions}
\label{sec:substellar}

Studies of substellar and planetary companions are generally focused on finding and characterizing companions orbiting FGK dwarf stars near the solar neighborhood.  As yet, no consensus exists regarding the demographics of substellar companions of evolved stars, including the companions' survivability as the stellar hosts evolve and their prevalence as a function of stellar metallicity \citep[like the metallicity-planet frequency trend seen in dwarf stars;][]{Fischer_2005_planetmetallicitycorr,Reffert_2015_evolvedstarplanets}.  

The companions of red giant stars are especially challenging to characterize because (1) the stellar host masses are difficult to constrain, and (2) both the planet detectability (influenced by the stellar RV jitter) and the system's architecture (influenced by star-planet tidal interactions and/or planetary engulfment) are expected to evolve as the star climbs the RGB. Therefore, to understand this class of substellar systems, the frequency of companions must be measured around a statistically large set of evolved stars that vary in age, composition, and galactic environment.

The primary objective for this class of APOGEE-2 targets is to increase the number of red giant stars known to have substellar companions, with particular attention to probing a range of stellar mass, metallicity, and age.  An additional benefit to this study is APOGEE-2's large set of chemical abundances for these hosts.  Previous studies of dwarf stars have indicated that planet-hosting stars have unique chemical abundance signatures \citep[e.g.,][]{Adibekyan_2012_planethostabundances}, some of which may be useful for constraining the planets' composition \citep[e.g.,][]{DelgadoMena_2010_planetabundances}; however, agreement between the reported signatures varies, and similar studies of the abundance patterns in evolved host stars have just begun to emerge \citep{Jofre_2015_evolvedhostabundances,Maldonado_2016_evolvedhostabundances}.

To maximize the temporal baseline available to characterize companion orbits, APOGEE-2's substellar companion search focuses on stars with a large number of RV measurements already taken with APOGEE-1.  Of the fields with $\geq$4 APOGEE-1 epochs, five were selected for follow-up in APOGEE-2 based on the number of epochs, position in the sky, and diversity of environment.  The outer disk fields {\it 120--08-RV}, {\it 150--08-RV}, and {\it 180--08-RV} probe stars in subsolar metallicity environments.  The open cluster NGC~188 (field {\it N188-RV}) was chosen to provide a sample for which stellar mass and age are known, and in which potential abundance signatures in host stars can be tested against non-host stars of similar chemistry.  The field {\it COROTA2-RV} includes stars with asteroseismically determined masses, which decreases the uncertainty in the mass of the companions.  The evolutionary stage information also allows discrimination between first-ascent red giant and red clump stars to study the process of tidal engulfment of planets on the RGB. 

Each of these ``{\it -RV}'' fields is observed numerous times to reach a final count of $\geq$24 epochs for each target.  The visits are cadenced such that the RV curves are sensitive to companions having a range of periods from a few days to nearly a decade (when combined with APOGEE-1 observations).  Within each field, the stars are selected from those targeted by APOGEE-1, prioritized first by the number of APOGEE-1 epochs and then by brightness, with brighter stars receiving higher priority.  Stars targeted as part of this class have the targeting flag APOGEE2\_TARGET3=4 set.


\subsection{RR Lyrae}
\label{sec:rrl}

Stellar distances larger than the reach of parallax measurements are notoriously difficult to measure accurately.  The discovery of the relationship in pulsational variable stars between stellar luminosity and the period of luminosity variation was a significant leap forward in understanding our very location in the Universe \citep[e.g.,][]{Leavitt_1912_cepheids}.  Now, precision multi-wavelength photometry of variable stars, especially RR Lyrae (RRL) and Cepheids, is enabling increasingly precise distance measurements to structures in the MW and Local Group \citep[e.g.,][]{Dekany_2013_bulgeRRL}; for example, individual RRL distance uncertainties of $\sim$1--2\%  are achievable with IR photometry \citep[e.g.,][]{Klein_2014_MIRLeavittLawWISE,Scowcroft_2015_CRRP,Beaton_2016_CCHP}.

APOGEE-2 is targeting $\sim$10,000 RRL stars, predominantly towards the inner Galaxy in the South.  A small number are also being observed with the 1-m and 2.5-m telescopes in the North.  High-resolution IR spectra of RRL with known proper motions confer some unique information: RVs and chemical abundances for old stars of known distance (thus 3-D space velocity), especially in dusty regions of the Galaxy that the Large Synoptic Survey Telescope, {\it Gaia}, and other optical facilities will not generally access \citep{Ivezic_2008_lsst,GaiaCollab_2016_gaia}.   The combination of precision distances and precision RVs provides powerful leverage for understanding the 6-D phase space of these obscured relics of the earlier Galaxy.

RRL observed by the 1-m are drawn from a sample of Northern hemisphere stars bright enough ($H \lesssim 10$) that single epochs yield reliable RV measurements.  These stars are used to build the ``RV templates'' --- relationships between pulsation phase and atmospheric velocity --- 
needed to correct single-epoch RV measurements to the true systemic velocity.  These new $H$-band relationships are needed because the RV templates that have been constructed for optical RV measurements \citep[e.g.,][]{Liu_1991_RRLRVtemplates,Sesar_2012_RRLRVtemplates} may not be applicable to $H$-band absorption lines, which probe different physical layers of the stars.

For APOGEE-2S, the RRL sample is drawn from the Optical Gravitational Lensing Experiment \citep[OGLE;][]{Udalski_2009_OGLE} catalog of variable stars \citep{Soszynski_2014_OGLERRL}, and generally spans $|l|\lesssim 11^\circ$ and $-8^\circ < b < -1^\circ$ or $1^\circ < b < 6^\circ$, following the OGLE footprint.  Proper motions are being calculated from the OGLE data, supplemented by {\it Gaia}. The simple selection criteria comprise (1) a magnitude limit of $H<14.5$, to reach a S/N sufficient for RV measurement, and (2) the requirement that the total flux from nearby stars cannot be greater than 15\% of the target flux (within 1.3$^{\prime\prime}$, roughly 2$\times$ the LCO fiber radius).  (We note that the RRL program may evolve, and users are encouraged to confirm these criteria in the relevant data release documentation.)  

All OGLE-classified RRL meeting these criteria are targeted for observation, during either normal APOGEE-2 survey time or during CIS-led time (\S\ref{sec:external}).  
These stars are either observed on all-RRL, single-visit plates, or on shared plates where unfilled fibers are placed on main sample red giant stars (\S\ref{sec:main_sample}) or other targets.  

All pre-selected RRL stars have the targeting bit APOGEE2\_TARGET1=24 set, potentially in addition to the 1-m target flag (APOGEE2\_TARGET2=22; \S\ref{sec:1m}) and others, if applicable.


\subsection{M Dwarfs}
\label{sec:m_dwarfs}

M~dwarf stars ($T_{\rm eff} \sim 2300-4000$~K) are highly valued for the study of planetary systems and stellar populations.  These stars are being targeted by numerous planet-hunting missions, including {\it K2} and TESS \citep{Ricker_2014_tess}, due to the smaller radius of their habitable zones and thus the relatively stronger signal of Earth-sized planets.  These long-lived, unevolved stars are also the most numerous stars in the Galaxy, making them useful tracers of the star formation and chemical enrichment history of the Galaxy's nearby stellar populations.  However, the densely lined and essentially continuum-less optical spectra of M dwarfs are notoriously difficult to analyze, and they are fainter at optical wavelengths than in the IR.  Thus, IR spectroscopy has become the state-of-the-art tool for measuring the dynamics and chemistry of these stellar tracers \citep[e.g.,][]{Onehag_2012_mdwarfsIR,Schmidt_2016_apogeeKMdwarfs,Souto_2017_apogeeMdwarfplanets}.

APOGEE-1 observed a substantial ancillary program of $\sim$1400 M~dwarfs to measure their RVs, RV variability, and rotational velocities \citep[see \S\,C.4 of Zasowski et al.\ 2013;][]{Deshpande_2013_apogeeMdwarfs}.  In APOGEE-2, we enhance this sample by targeting $\sim$5000 known M~dwarfs in the MaNGA-led halo fields (\S\ref{sec:main_sample}). These M~dwarfs were drawn from multiple catalogs \citep[e.g.,][]{Reid_2005_faintPMstars,Lepine_2005_LSPMNorth,Lepine_2011_Mdwarfcatalog,Gaidos_2014_coolstarcatalog}, within a magnitude range of $7 < H < 12$ and with a color requirement of $(V-J)>0$.  Within the APOGEE-2 targets on each MaNGA-led design, an average of seven M~dwarfs are targeted and prioritized by $(V-J)$ color.
In addition, some M~dwarfs in the {\it Kepler} footprint are targeted as part of the APOGEE-2 ancillary program. 
These M~dwarf targets are flagged with the targeting bit APOGEE2\_TARGET3=3 and/or any relevant ancillary program bits.


\subsection{Eclipsing Binaries}
\label{sec:ebs}

Eclipsing binary (EB) systems contain at least two stars whose orbits lie in a plane nearly parallel to the line of sight, and thus the stars undergo periodic mutual eclipses.  EBs have long been used as laboratories with which to study the fundamental mass-radius relationship of stars \citep{Torres_2010_EBreview}.  Most often discovered through photometric variability, EBs usually require spectroscopic follow-up to determine their orbital velocities and fundamental stellar parameters.  One benefit of observing these in the IR is that reliable RVs can be measured for 
systems with favorable flux contrast ratios, such as those with low mass secondaries (e.g., M~dwarfs; \S\ref{sec:m_dwarfs}) orbiting solar-like stars.

Approximately 100 EBs were targeted in APOGEE-1, predominantly in the {\it Kepler} footprint.  In APOGEE-2, this sample is roughly tripled to include additional {\it Kepler}-detected EBs as well as systems identified in the Kilodegree Extremely Little Telescope survey \citep[KELT;][]{Pepper_2007_KELT,Pepper_2012_KELTsouth}, which spans nearly the entire sky.  

The {\it Kepler} EBs are hand-selected to focus on the most
astrophysically-rich systems: those with very shallow secondary eclipse depths,
evidence of third-body interactions, very long orbital periods, or out-of-eclipse variations that may be caused by pulsations.  The {\it Kepler} targets
are selected from the {\it Kepler} EB Catalog’s list of detached EBs
\citep{Prsa_2011_keplerEBs,Slawson_2011_keplerEBs}, using a magnitude limit of $H \leq 13$.  Up to ten EB targets are selected in each APOGEE-2 {\it Kepler} pointing for the EB program.

The KELT-based sample is selected from systems lying
in already-planned APOGEE-2 field locations that are anticipated to be observed for eight epochs over the course of the survey. KELT itself is restricted to bright stars ($V<10$), so no additional magnitude cuts are required. 
KELT targets are selected with very minimal criteria: a well-defined orbital period, with further preference towards those systems that have a detached morphology, are bright, and/or have shallow secondary eclipses.  Up to a maximum of five KELT targets are allocated for the fields that include KELT EBs; for most fields there are fewer than five KELT EB targets, so it is rare that anything other than the orbital period and binary nature of the systems are used to assign targets.

All EB targets, both {\it Kepler}- and KELT-selected, are flagged with the APOGEE2\_TARGET3=1 targeting bit.


\subsection{Bar/Bulge Red Clump Giants}
\label{sec:bulge_rc}

Red clump (RC) giants are metal-rich, horizontal branch (He burning) stars.  They span a very small range of absolute magnitude and color \citep[e.g.,][]{Girardi_2001_RCabsmag}, thus making them invaluable as standard candles (and standard crayons) to meausure stellar distances and foreground dust reddening \citep[e.g.,][]{Stanek_1997_RCbar,Benjamin_05_glimpse,Nataf_2013_bulgeextlaw}.  Because RC stars meet APOGEE's main sample color criteria (\S\ref{sec:main_sample}), many stars in the disk and halo sample are RC giants; \citet{Bovy_2014_APOGEE_RC_catalog} compiled a catalog of these stars for general use.

Unfortunately, the magnitude of the central bar/bulge's RC population ($H \sim 14$) is fainter than APOGEE-2's typical limits for red giants.  But because of the value of these stars for tracing the chemodynamics of distance-resolved structures, such as the bar and X-shape \citep[e.g.,][]{McWilliam_10_xshapedbulge,Nataf_2010_dualbulgeRC,Wegg_2013_RCbulge}, APOGEE-2 targets a few ``deep'' inner Galaxy fields, with a targeting strategy designed to increase the fraction of RC to RGB stars.  

Along the bulge's minor axis ($l=0^\circ$), fields at $b=\pm 8^\circ$ and $\pm12^\circ$ are planned for up to 18 visits.  The candidate RC stars in the $b>0^\circ$ fields are selected in a customized range of dereddened color: $(J-K_s)_0 \geq 0.745$ for $(0^\circ,8^\circ)$ and $(J-K_s)_0 \geq 0.52$ for $(0^\circ,12^\circ)$; results from these fields will inform the selection in the $b<0^\circ$ fields for further optimization of the RC yield. The magnitude range is chosen to bracket the visible number count ``peaks'' in  WISE 4.5$\mu$m and de-reddened $K_s$ photometry (associated with the RC stellar density), following \citet{Benjamin_05_glimpse}, \citet{Zasowski_2012_innerMW}, and \citet{Wegg_2013_RCbulge}.  Stars within these color and magnitude ranges are randomly selected in order to sample the full RC (and RGB contaminant) distribution between the peaks.

These bar/bulge RC candidates have the targeting bit APOGEE2\_TARGET1=25 set.


\subsection{Ancillary Programs}
\label{sec:ancillary}

Two calls for APOGEE-2 ancillary proposals were issued, resulting in 23 programs.
All targets observed as part of an ancillary program have bit ${\rm APOGEE2\_TARGET}3={\rm 8}$ set,
along with the ${\rm APOGEE2\_TARGET}3$ bit for specific programs (Table~\ref{tab:targeting_bits}).
Targeting and other information for each program will be included in data releases containing those data.

\subsection{1-m Targets}
\label{sec:1m}

The APOGEE2-N spectrograph has a single fiber connection to the NMSU 1-meter telescope located next to the Sloan 2.5-meter at APO \citep{Holtzman_2010_nmsu1m,Majewski_2017_apogeeoverview}.  When the spectrograph is not employed in observations on the 2.5-m, the 1-m telescope can be used to observe other high-priority targets.  These include (1) very bright stars that would saturate during a standard APOGEE visit, such as stars with high-resolution optical measurements or sparsely distributed ancillary program targets, and (2) isolated stars needing numerous epochs, as for the construction of RV template curves for RR Lyrae (\S\ref{sec:rrl}).  

Observations taken with the 1-m telescope are flagged with the APOGEE2\_TARGET2=22 bit.  These data may also have other targeting bits set, but the ``field name'' associated with each source identifies the targeting type or program.

\subsection{External Programs}
\label{sec:external}

The APOGEE2-S spectrograph is also used for observations during time allocated independently by the Carnegie Institution of Science (CIS) and the Chilean National Time Allocation Committee (CNTAC).  The targeting of sources observed during this time is completely independent from any of the APOGEE-2 procedures outlined in this paper, beyond some basic restrictions so that the same scheduling software can be used.  Targets observed during CIS-allocated time are flagged with APOGEE2\_TARGET2=24, and those during CNTAC-allocated time are flagged with APOGEE2\_TARGET2=25.

The PIs of these ``external'' programs choose whether the observed data are included in the SDSS Data Releases.  Description of the external programs that are contributed to SDSS will be included in the relevant data release documentation.  Programs that are non-contributed remain proprietary to the team allocated time by CIS or the CNTAC.  These proprietary targets have bit APOGEE2\_TARGET2=26 set; by definition, these observations are not intended to appear in the SDSS data releases, but we include the flag here for completeness and in case of future changes.

\section{Calibrator Target Selection}
\label{sec:calibration_targets}

APOGEE-2 has two types of targets called ``calibrators'': (1) observations used to correct the science data (\S\ref{sec:calibration_observation}) and (2) observations used to calibrate the derived stellar parameters and abundances with those of other studies (\S\ref{sec:calibration_params}-\ref{sec:calibration_surveys}).

\subsection{Telluric Absorption and Sky Emission Calibration}
\label{sec:calibration_observation}

As in APOGEE-1, atmospheric H$_2$O, CO$_2$, and CH$_4$ contribute substantial absorption features to every observed APOGEE-2 spectrum.  To separate these lines from the stellar and interstellar features, and perform telluric corrections to the observed spectra, APOGEE-2 continues APOGEE-1's procedure of observing several early-type stars in every field \citep[\S5.1 of][]{Zasowski_2013_apogeetargeting}.  

Between 15 and 35 of the bluest stars in the field are targeted in every design.  The design's FOV is divided into several equal-area zones; half of the ``telluric calibrator'' stars are selected as the bluest star in their zone, and the other half are the bluest stars remaining anywhere in the FOV.  This method ensures that the telluric calibrators will be among the earliest-type stars in the field, but also that they are spread somewhat evenly over the FOV.  This latter criterion is critical, as the APO and LCO FOVs are large enough that the strength of the telluric absorption lines can vary from target to target.  See \S5.1 of \citet{Zasowski_2013_apogeetargeting} for further details.

Telluric calibrator targets are prioritized over all other targets in a design, and can be identified by the targeting bit APOGEE2\_TARGET2=9.

In addition to telluric absorption, the Earth's atmosphere contributes substantial $H$-band {\it emission} via IR airglow lines (primarily from OH) and scattered light from the Moon; additional background is contributed by zodiacal dust and unresolved stars.  APOGEE-2 uses the same strategy adopted in APOGEE-1 of observing several representative ``empty'' sky positions on each design, to cleanly measure the contaminating emission.  The selection procedure is described more fully in \S5.2 of \citet{Zasowski_2013_apogeetargeting}, but in essence, $\sim$35 positions that are devoid of 2MASS sources within the 6$^{\prime\prime}$ neighbor limit applied to the main sample sources are targeted as part of every design.  These 35 positions are selected from the same areal zones used in the selection of the telluric calibrators (above), ensuring an even sampling across the FOV.

These ``sky'' targets are prioritized after all other targets in a design, and can be identified with the targeting bit APOGEE2\_TARGET2=4.

\subsection{Stellar Parameters and Abundances Calibration}
\label{sec:calibration_params}

A subset of stars in many of the targeted open and globular clusters have existing multi-element abundance, stellar parameter, and radial velocity derivations from high resolution optical spectroscopy, in many cases homogeneously observed and analyzed \citep[e.g.,][]{carrettauves,carrettagiraffe}.  These stars have the targeting bits APOGEE2\_TARGET2=2 and APOGEE2\_TARGET2=10 set, and allow a direct comparison of elemental abundances and stellar parameters derived by ASPCAP against those derived using optical high-resolution spectra \citep[e.g.,][]{Meszaros_2013_aspcapcalib,Holtzman_2015_apogeedata}.  Literature abundances were used to ensure that these targets span the critical metallicity and abundance ranges.  For example, the globular clusters span the known internal cluster abundance anticorrelations in certain elements (e.g., Mg--Al, Na--O), as well as the full observed range of [Fe/H] in the clusters where a spread in [Fe/H] has been observed \citep[e.g.,][]{m22_iron,omegacen_sandro}.     

\subsection{Cross- and Inter-Survey Calibration}
\label{sec:calibration_surveys}

At least four fields are being observed from both the North and the South with as many stars in common as possible, allowing a direct comparison of instrument performance and derived abundances and stellar parameters between the two spectrographs.  These fields include the globular cluster M12, two open clusters (M67 and NGC~2243), and one field at ($l$,$b$)=(0$^\circ$,8$^\circ$).

APOGEE-2 targets overlap, to various extents, with numerous ongoing photometric and spectroscopic surveys.  These cases provide valuable opportunities to enhance our knowledge of stellar astrophysics by leveraging APOGEE-2 spectra together with complementary observations exploring other wavelength regimes and/or observational sampling.  This is exemplified by recent results from the APOKASC \citep{tayar_mixlen} and CoRoGEE \citep{Anders_2017_corotgee} samples, and a plethora of additional opportunities lie with optical spectroscopic surveys such as GALAH \citep{galah} and GES \citep{gilmore_ges}.  Targets selected to be in common with other datasets, spectroscopic and photometric, have the targeting bit APOGEE2\_TARGET2=5 set, as well as bits for specific surveys (Table~\ref{tab:targeting_bits}).

\section{Targeting Information in Data Releases}
\label{sec:dr_targeting_info}

In addition to the spectra and derived stellar parameters, 
APOGEE data releases contain the pre-targeting and selection information necessary to understand the selection function of the sample.
Along with the targeting flags described in \S\ref{sec:targeting_flags}, 
this information is contained within four types of files:
\begin{description} \itemsep -2pt
\item[apogee2Object] ID, coordinates, photometry, proper motions, etc.\ for each {\it object} within all survey field footprints
\item[apogee2Field] central ($\alpha, \delta$), location ID, field name, planned number of visits for each {\it field} in the survey
\item[apogee2Design] design ID, central ($\alpha, \delta$), location ID, radius, cohort versions, cohort \& calibration fiber allocation, cohort magnitude ranges, color range, planned number of visits for each {\it design} in the survey
\item[apogee2Plate] plate ID, design ID, location ID, drill angle, drill temperature, drill epoch for each {\it plate} in the survey
\end{description}

APOGEE-2 is providing a unique window into the workings of the MW, at a level of detail unobservable in any other large galaxy.  This expansive, homogeneously analyzed, all-sky dataset will enable significant leaps forward in our understanding of star formation, stellar system architecture, stellar astrophysics, and galaxy formation and evolution on all scales.

\begin{acknowledgments}

G.Z.~acknowledges support from the Barry~M.~Lasker Data Science Research Fellowship, sponsored by the Space Telescope Science Institute in Baltimore, MD, USA.
R.E.C.~acknowledges funding from Gemini-CONICYT for Project 32140007, 
C.B.~acknowledges support from National Science Foundation (NSF) grant AST-1517592, and
P.M.F.~acknowledges support from NSF grant AST-1311835 and AST-1715662.
D.A.G.H was funded by the Ram\'{o}n y Cajal fellowship number RYC-2013-14182, and D.A.G.H and
F.D. acknowledge support provided by the Spanish Ministry of Economy and
Competitiveness (MINECO) under grant AYA-2014-58082-P.
Sz.M. has been supported by the Premium Postdoctoral Research Program of the Hungarian Academy of Sciences, and by the Hungarian
NKFI Grants K-119517 of the Hungarian National Research, Development and Innovation Office.
We thank P.~Stetson for access to optical photometry that improved the selection of globular cluster targets.
Finally, we thank the anonymous referee for comments that improved the clarity of this paper.

Funding for the Sloan Digital Sky Survey IV has been provided by the Alfred P. Sloan Foundation, the U.S. Department of Energy Office of Science, and the Participating Institutions. SDSS-IV acknowledges support and resources from the Center for High-Performance Computing at the University of Utah. The SDSS web site is www.sdss.org.

SDSS-IV is managed by the Astrophysical Research Consortium for the Participating Institutions of the SDSS Collaboration including the Brazilian Participation Group, the Carnegie Institution for Science, Carnegie Mellon University, the Chilean Participation Group, the French Participation Group, Harvard-Smithsonian Center for Astrophysics, Instituto de Astrof\'isica de Canarias, The Johns Hopkins University, Kavli Institute for the Physics and Mathematics of the Universe (IPMU) / University of Tokyo, Lawrence Berkeley National Laboratory, Leibniz Institut f\"ur Astrophysik Potsdam (AIP),  Max-Planck-Institut f\"ur Astronomie (MPIA Heidelberg), Max-Planck-Institut f\"ur Astrophysik (MPA Garching), Max-Planck-Institut f\"ur Extraterrestrische Physik (MPE), National Astronomical Observatory of China, New Mexico State University, New York University, University of Notre Dame, Observat\'ario Nacional / MCTI, The Ohio State University, Pennsylvania State University, Shanghai Astronomical Observatory, United Kingdom Participation Group, Universidad Nacional Aut\'onoma de M\'exico, University of Arizona, University of Colorado Boulder, University of Oxford, University of Portsmouth, University of Utah, University of Virginia, University of Washington, University of Wisconsin, Vanderbilt University, and Yale University.

This publication makes use of data products from the Two Micron All Sky Survey, which is a joint project of the University of Massachusetts and the Infrared Processing and Analysis Center/California Institute of Technology, funded by the National Aeronautics and Space Administration and the National Science Foundation.  This publication also makes use of data products from the Wide-field Infrared Survey Explorer, which is a joint project of the University of California, Los Angeles, and the Jet Propulsion Laboratory/California Institute of Technology, funded by the National Aeronautics and Space Administration.

\end{acknowledgments}

\bibliographystyle{apj}
\bibliography{apogee2_targeting.bib}

\begin{appendix}

\section{Glossary}
\label{sec:glossary}

This Glossary contains SDSS- and APOGEE-specific terminology appearing in this paper and throughout the data documentation.

\begin{description} \itemsep -2pt
\item[1-Meter Target] Target observed with the NMSU 1-m telescope, which has a single fiber connection to the APOGEE-2N instrument (\S\ref{sec:1m}).
\item[Ancillary Target] Target observed as part of an approved ancillary program.
\item[ASPCAP] The APOGEE Stellar Parameters and Chemical Abundances Pipeline; the analysis software that calculates basic stellar parameters (T$_{\rm eff}$, $\log{g}$, [Fe/H], [$\alpha$/Fe], [C/Fe], [N/Fe]) and elemental abundances \citep{Holtzman_2015_apogeedata,GarciaPerez_2016_aspcap}.
\item[Cohort] Set of targets in the same field that are observed together on all of their visits (\S\ref{sec:design_defs}).  A given plate may have multiple cohorts on it.
\item[Design] Set of targets drilled together on a plate, consisting of up to one each of short, medium, and long cohorts (\S\ref{sec:design_defs}).  A design is identified by an integer Design ID.  
\item[Design ID] Unique integer assigned to each design.
\item[Drill Angle] Hour angle (distance from the meridian) at which a plate is drilled to be observed.  This places the fiber holes in a way that accounts for differential refraction across the FOV.
\item[External Program] General term for programs and targets observed during the APOGEE-2S time allocated by the Carnegie Observatories (OCIS) or the Chilean Time Allocation Committee (CNTAC) (\S\ref{sec:external}).  These targets may or may not be included in the SDSS dataset.
\item[Fiber Collision] A situation in which two targets, separated by less than the protective ferrule around the fibers, are included in the same design.  The higher-priority target will be drilled onto the plate(s); the lower-priority target will be removed.
\item[Fiber ID] Integer (1--300) corresponding to the row on the detector from which the spectrum was extracted.  Fiber IDs can vary from visit to visit for a given star.
\item[Field] Location on the sky, defined by central coordinates and a radius (\S\ref{sec:field_properties}.  
\item[Location ID] Unique integer assigned to each field on the sky.
\item[Plate] Piece of aluminum with a design drilled on it.  Note that while ``plate'' is often used interchangeably with ``design'', multiple plates may exist for the same design -- e.g., drilled at different hour angles.
\item[Plate ID] Unique integer assigned to each plate.
\item[RJCE] The Rayleigh-Jeans Color Excess method, a technique used to estimate the line-of-sight reddening to a star.  APOGEE-2 uses this method to estimate intrinisic colors for many potential targets.
\item[Sky Targets] Empty regions of sky observed on a plate in order to remove the atmospheric airglow lines and sky background from the target spectra.
\item[Special Targets] General term for targets selected with criteria other than the color and magnitude criteria of the main red giant sample (\S\ref{sec:main_sample}).  For example, special targets include ancillary program targets and calibration cluster members.
\item[Targeting Flag and Bits] A targeting ``flag'' refers to one of the three long integers assigned to every target in a design, each made up of 31 ``bits'' that correspond to particular selection or assignment criteria (\S\ref{sec:targeting_flags}).  APOGEE-2's flags are named APOGEE2\_TARGET1, APOGEE2\_TARGET2, and APOGEE2\_TARGET3; see Table~\ref{tab:targeting_bits} for a list of the bits.
\item[Telluric Standards] Hot blue stars observed on a plate to derive corrections for the telluric absorption lines (\S\ref{sec:calibration_observation}).
\item[Visit] The base unit of observation, equivalent to approximately one hour of on-sky integration (but this can vary) and comprising a single epoch.  Repeated visits are used to both build up signal and provide a measure of RV stability.
\item[Washington + DDO51] Also ``W+D photometry''; adopted abbreviation for the combination of Washington $M$ and $T_2$ photometry with {\it DDO51} photometry, used in the classification of dwarf/giant stars.
\end{description}

\end{appendix}

\end{document}